\documentclass[aps,showpacs,showkeys,onecolumn,floatfix,nofootinbib]{revtex4}
\usepackage{graphicx,epsfig,wrapfig,color}
\usepackage{amsmath}
\usepackage{amssymb}
\usepackage{dsfont}
\usepackage{color}

\newcommand{\be}{\begin{equation}}
\newcommand{\ee}{\end{equation}}
\newcommand{\ba}{\begin{eqnarray}}
\newcommand{\ea}{\end{eqnarray}}

\newcommand{\di}{\!{\rm d}}
\newcommand{\la}{\langle}
\newcommand{\ra}{\rangle}

\begin{document}
\newcommand*{\UConn}{
   Department of Physics, University of Connecticut,
   Storrs, CT 06269-3046, U.S.A.}\affiliation{\UConn}
\title{\boldmath
        Effects of long-range forces on the $D$-term and the energy-momentum structure}
\author{Mira Varma}\author{Peter Schweitzer}\affiliation{\UConn}
 \date{June 2020}
\begin{abstract}
The hadronic form factors of the energy-momentum tensor (EMT)
have attracted considerable interest in recent literature.
This concerns especially the $D$-term form factor $D(t)$ with 
its appealing interpretation in terms of internal forces. 
With their focus on hadron structure, theoretical studies so far 
have concentrated on strongly interacting systems with short-range forces. 
Effects on the EMT due to long-range forces like the electromagnetic 
interaction have not yet been studied. Electromagnetic forces play a 
small role in the balance of forces inside the proton, but their 
long-range nature introduces new features which are not present 
in systems with short-range forces. 
We use a simple but consistent classical field theoretical model of 
the proton to show how the presence of long-range forces alters some 
notions taken for granted in short-range systems. Our results imply 
that a more careful definition of the $D$-term is required when 
long-range forces are present. 
\end{abstract}
%
%
%
%
\maketitle

\section{Introduction}

The matrix elements the EMT, $T_{\mu\nu}$, \cite{Kobzarev:1962wt} can be
explored through studies of generalized parton distribution functions 
in hard exclusive reactions \cite{Mueller:1998fv,Ji:1998pc} and contain 
information on the basic properties of a particle: mass, spin, and the 
equally important but far less known $D$-term \cite{Polyakov:1999gs}.
The information content of EMT form factors is visualized in terms of EMT 
densities \cite{Polyakov:2002yz} which allow us to learn about properties 
like energy density, angular momentum distribution, or internal forces in 
hadrons \cite{Polyakov:2002yz,Lorce:2017wkb,Polyakov:2018guq,
Polyakov:2018zvc,Polyakov:2018exb,Lorce:2018egm,Schweitzer:2019kkd}.
EMT properties were studied in hadronic models, chiral perturbation 
theory, lattice QCD and other strongly inter\-acting systems 
\cite{Ji:1997gm,Petrov:1998kf,Schweitzer:2002nm,Ossmann:2004bp,
Goeke:2007fp,Goeke:2007fq,Wakamatsu:2007uc,
Cebulla:2007ei,Jung:2013bya,Kim:2012ts,Jung:2014jja,
Mai:2012yc,Mai:2012cx,Cantara:2015sna,Gulamov:2015fya,Nugaev:2019vru,
Donoghue:1991qv,Kubis:1999db,Belitsky:2002jp,Ando:2006sk,Diehl:2006ya,
Hagler:2003jd,Gockeler:2003jfa,Hagler:2007xi,
Shanahan:2018nnv,Pasquini:2014vua,
Grigoryan:2007vg,Pasquini:2007xz,Hwang:2007tb,Abidin:2008hn,
Brodsky:2008pf,Chakrabarti:2015lba,
Kumar:2017dbf,Mondal:2017lph,Hudson:2017xug,Hudson:2017oul,Anikin:2019kwi,
Neubelt:2019sou,Azizi:2019ytx,Ozdem:2019pkg}
which had one common feature: 
these systems were governed by {\it short-range} forces. 
The goal of this work is to investigate the impact of {\it long-range}
forces on the EMT properties. 

For our study, we employ a classical model of the proton which is of 
interest for its own sake. Classical models of an extended electric 
charge have a long history dating back to the works of Abraham and 
Lorentz \cite{Abraham,Lorentz}. It was recognized by Poincar\'e that 
in order to compensate the electrostatic repulsion one must introduce 
cohesive forces, known as Poincar\'e stresses \cite{Poincare}, which 
were introduced in an {\it ad hoc} manner \cite{Abraham,Lorentz,Poincare, 
Dirac:1962iy,Schwinger:1983nt,BialynickiBirula:1984ei,Pearle}.
The model of Bia\l ynicki-Birula \cite{BialynickiBirula:1993ce} 
used in this work is to the best of our knowledge the 
first fully consistent classical model of an extended charged particle 
where the Poincar\'e stresses are generated dynamically in a local, 
relativistic, classical field theory.

In this model, ``dust particles'' carry an electric charge $e$, 
and strong charges $g_S$ and $g_V$ and interact with the electromagnetic 
4-potential $A^\mu$, and strong scalar and vector fields, $\phi$ and $V^\mu$.
The attractive (due to $\phi$) and repulsive (due to $V^\mu$ and $A^\mu$) 
forces on the dust particles exactly compensate each other, such that the 
dust particles are in stable, static equilibrium and occupy a finite 
spherically symmetric region of radius $R$.
Using nuclear phenomenology to fix model parameters, the model can 
describe a particle with the charge, mass, and size of the proton
\cite{BialynickiBirula:1993ce}.

This classical system is well-suited for our purposes. It exhibits 
strong short-range forces which play an overwhelmingly important 
role in the internal structure of the proton, and at the same time
 consistently includes the effects of the long-range electromagnetic 
field. 
Two aspects are of importance for our study, namely (i) an internally 
consistent theoretical description of a stable particle, and (ii) the 
correct description of the long-range electromagnetic effects.
The model of Ref.~\cite{BialynickiBirula:1993ce} satisfies both
requirements. The classical aspect of the model is not a hindrance. 
Rather it is a virtue allowing us to investigate the effects of long-range 
forces undistracted by technical difficulties associated with computations 
in more realistic strongly interacting quantum systems. 

The outline of this work is as follows: 
In Sec.~\ref{Sec-2:model} we briefly introduce the model, and apply 
it to the description of EMT densities in Sec.~\ref{Sec-3:EMT-in-model}, 
showing that the model is consistent and, in the region of
$r\lesssim 2\,{\rm fm}$, in good agreement with results
from studies of systems with strong short-range forces.
In Sec.~\ref{Sec-4:EMT+long-range-forces} we focus on distances beyond 
$r\gtrsim 2\,{\rm fm}$ where new features appear which were not encountered 
before in systems with short-range forces. 
In Sec.~\ref{Sec-5:model-independent-insights} we show that our
results regarding the long-distance properties of the EMT densities are model 
independent, compute the form factor $D(t)$, and discuss the implications
for experimental measurements and theoretical calculations of the $D$-term.
The Sec.~\ref{Sec-6:conclusions} contains the conclusions.

\newpage
\section{The Classical Model}
\label{Sec-2:model}

In this section, we briefly introduce the model of 
Ref.~\cite{BialynickiBirula:1993ce} which consists of 
``dust particles'' bound in a spherically symmetric region of 
radius $R$ by the interplay of three types of fields: a massive 
scalar field $\phi$, a massive vector field $V^\mu$ and 
an electromagnetic field described by the 4-potential $A_\mu$.
The particles couple to these fields respectively through
the coupling constants $g_S$, $g_V$, $e$. The motion of the 
particles is described by a scalar phase-space distribution 
$\Gamma(\vec{r}, \vec{p}, t)$. 
The system is defined by the following classical field equations
(we use $\hbar=c=1$ unless otherwise stated)
\ba
         \label{eq:classicalzero}
         \left[  ( m-g_{S} \phi  )  
           (\partial _{t}+\vec{v} \cdot \vec{\nabla}_{r}  ) 
           + m \, \vec{F} \cdot  \vec{\nabla}_{p} \right] 
           \Gamma(\vec{r},\vec{p},t) 
         & = & 0 \, , \\
         \label{eq:classicalthree}
	 \partial _{ \alpha }G^{ \alpha \beta}+m_{V}^{2}V^{\beta} 
         & = & g_{V} \, j^{\beta} ,\\
         \label{equation:classicaltwo}
	 ( \Box+ m_{S}^{2} )  \phi  
         & = & g_{S} \, \rho \, , \phantom{ \biggl|}\\
         \label{eq:classicalone}
	 \partial _{ \alpha }F^{ \alpha \beta} 
         & = & e \, j^{\beta} .
\ea
The force \( \vec{F}=\vec{f}/{u^{0}} \) 
is expressed in terms of the components of the 4-force
$f^{ \alpha } = eF^{ \alpha \beta}u_{\beta} +g_V G^{ \alpha \beta}u_{\beta}-g_{S} (  \partial ^{ \alpha }-u^{ \alpha }u^{\beta} \partial _{\beta} )  \phi$
with
\( F^{ \alpha \beta} =\partial^\alpha A^\beta-\partial^\beta A^\alpha\)  and  
\( G^{ \alpha \beta} =\partial^\alpha V^\beta-\partial^\beta V^\alpha\).
The 4-velocity $u^\alpha$ is defined by 
\( p^\alpha = m\, u^{ \alpha } = (E_p,\vec{p})\) and 
\( \vec{v}= \vec{u}/u^{0}=\vec{p}/{E_{p}} \).
\( \nabla_{r}  \)  and  \(  \nabla _{p} \)  denote derivatives 
with respect to positions $\vec{r}$ and momenta $\vec{p}$
of the particles. The 4-current \( j^{ \alpha } \) and scalar 
density  \(  \rho  \) are defined in terms of the phase-space 
distribution as
\ba
	j^{ \alpha }(\vec{r},t) 
        &=& \int\frac{d^{3}p}{E_{p}}\,p^{\alpha}\,\Gamma(\vec{r},\vec{p},t),\\
	\rho(\vec{r},t) 
        &=& \int\frac{d^{3}p}{E_{p}}\:m\:\Gamma(\vec{r},\vec{p},t).
\ea
Despite the non-covariant appearance of Eq.~(\ref{eq:classicalzero}) the 
theory described by Eqs.~(\ref{eq:classicalzero}--\ref{eq:classicalone}) 
is relativistically invariant \cite{BialynickiBirula:1993ce}, and is a
generalization of the Vlasov-Maxwell equations used in plasma physics 
\cite{Maxwell-Vlasov}. 

The solution of Eqs.~(\ref{eq:classicalzero}--\ref{eq:classicalone}) is 
most conveniently expressed in the static case, where the particles
are at rest with $u^\alpha=(1,0,0,0)$ and described by the phase space 
distribution $\Gamma(\vec{r},\vec{p},t) = \delta^{(3)}(\vec{p}) \:\rho(r)$ 
with $r = |\vec{r}|$. In this frame, the scalar density $\rho$ 
and zeroth component of  \( j^{ \alpha }\) coincide, $A^\alpha$ 
and $V^\alpha$ only have zero components, and 
Eqs.~(\ref{eq:classicalzero}--\ref{eq:classicalone}) become
\ba
        \label{eq:force-equilibrium}
	\rho \vec{F} \equiv - \rho \vec{\nabla}  
        ( eA_{0}-g_{S} \phi +g_V V_0 ) 
        & = & 0\,,\\
      \label{eq:minusdelmw}
	( - \Delta +m_V^{2} ) V_0
        & = & g_V  \rho \, , \\
        \label{eq:minusdelmphi}
	( - \Delta +m_{S}^{2} )  \phi 
        & = & g_{S} \rho \, , \\
        \label{eq:minusdelazero}
	- \Delta A_{0} 
        & = & e \rho \, .
\ea
Notice that the condition (\ref{eq:force-equilibrium}) provides a
constraint on the fields only for $r\le R$ where matter is present 
(i.e.\ $\rho\neq0$), and is trivially satisfied in the region $r>R$ 
with no matter (where $\rho=0$).
The density is normalized as $\int  d^{3}r \;\rho(r) = 1$.
In the region \( r \leq R \) the solutions of 
Eqs.~(\ref{eq:classicalzero}--\ref{eq:classicalone}) are given by
\cite{BialynickiBirula:1993ce}
\ba
         \label{eq:fplusfminus}
	 \rho(r) &=& f_{+}(r)-f_{-}(r) \phantom{\frac11}\\
         \label{eq:eazero}
	 eA_{0}(r) &=& e^{2}\biggl( 
         \frac{f_{+}(r)}{k_{+}^{2}}-\frac{f_{-}(r)}{k_{-}^{2}} \biggr) + 2E_B \\
         \label{eq:gsphi}
	 g_{S} \phi(r) &=& g_{S}^{2} \biggl(
         \frac{f_{+}(r)}{k_{+}^{2}+m_{S}^{2}}-\frac{f_{-}(r)}{k_{-}^{2}+m_{S}^{2}} 
         \biggr) \\
         \label{eq:gswzero}
	 g_V V_{0}(r) &=& g_V^{2} \biggl( 
         \frac{f_{+}(r)}{k_{+}^{2}+m_V^{2}}-
         \frac{f_{-}(r)}{k_{-}^{2}+m_V^{2}} \biggr) {.}
\ea
The functions $	f_{ \pm }(r)$ are defined by
\be
	f_{ \pm }(r)=\frac{d_{ \pm }}{4 \pi }\frac{\sin ( k_{ \pm }r ) }{r}{,}
        \quad 
	k_{ \pm }^{2}=\frac{B \pm \sqrt{D}}{2Q^2}{,}
\ee
where 
$B = (g_{S}^{2}-e^{2}) m_V^{2} - ( g_V ^{2}+e^{2} ) m_{S}^{2}$ and 
$D = B^2 - 4 e^2 Q^2 m_S^2 m_V^2$ with
$Q^2 = e^{2}-g_{S}^{2}+g_V ^{2}$ (notice the misprint in 
Eq.~(23) of \cite{BialynickiBirula:1993ce} in the definition of
$k_\pm$).

In the region \( r>R \) the solutions of 
Eqs.~(\ref{eq:classicalzero}--\ref{eq:classicalone}) are given by
\ba
         \label{eq:rhozero}
	 \rho(r) &=& 0 , \\
         \label{eq:eazero-out}
	 eA_{0}(r) &=& \frac{e^{2}}{4 \pi r} \\
         \label{eq:gsphisolution}
	 g_{S} \phi(r) &=& \frac{b_{S}}{4 \pi r} e^{-m_{S} ( r-R ) } \\
         \label{eq:gnuwsolution}
	 g_V V_{0}(r) &=& \frac{b_V}{4 \pi r}e^{-m_V ( r-R ) }\,.
\ea
Eq.~\eqref{eq:rhozero} means there are no particles outside
the radius $R$, and Eq.~\eqref{eq:eazero-out} is the Coulomb potential.
The six parameters  \( b_V,\;b_{S},\;d_{+},\;d_{-},\;2E_B,\;R\) are 
fixed by requiring the fields \( A_{0}(r),\;V_0(r), \;\phi(r) \) 
to be continuous and differentiable at \(r=R\). The reason why the
constant $2E_B$ has been named in this peculiar way will become
clear shortly.

In order to apply the model to the description of the proton the
following parameters were used in Ref.~\cite{BialynickiBirula:1993ce} 
\ba
        m   = 938 \,{\rm MeV}, \quad
	m_S = 550 \,\frac{\rm MeV}{\hslash c}, \quad 
	m_V = 783 \,\frac{\rm MeV}{\hslash c}, \quad 
	\frac{g_{S}^{2}}{\hslash c} = 91.64 \, , \quad
	\frac{g_V^{2}}{\hslash c} = 136.2 \, , \quad 
    \alpha = \frac{e^2}{4\pi\hslash c} = \frac{1}{137}, \, 
    \label{eq:parameters} 
\ea
where for convenience the constants $\hbar c = 197\,{\rm MeV\,fm}$ 
are restored such that in all expressions, $r$ is in units of fm, 
energies in units of MeV, $\rho(r)$ in units of fm$^{-3}$, etc. The
parameters $m_S$ and $m_V$ correspond respectively to the masses of a 
$\sigma$-meson and $\omega$-meson as used in nuclear matter models. 
The coupling constants $g_S$, $g_V$ are taken from the model QHD-I 
of the mean field theory of nuclear matter \cite{nuclear-matter}. 
This means that in this model the proton is bound by nuclear forces 
\cite{BialynickiBirula:1993ce}. 
For completeness, we remark that with these parameters, 
the requirements of continuity and differentiability 
of $\phi(r), A_0(r), V_0(r)$ at $r=R$ fix the constants 
$b_V,\,b_s,\,d_\pm ,\,2E_B,\,R$ to have the following values:
$b_V=1354.13\,{\rm MeV\,fm}$,
$b_{S}=1786.38\,{\rm MeV\,fm}$,
$d_{+}=2.02477 /{\rm fm}^2$,
$d_{-}=-3.93639 /{\rm fm}^2$,
$2E_B=-31.42 \,{\rm MeV}$,
$R=1.05 \, {\rm fm}$.

The parameter $E_B = -15.71\,{\rm MeV}$ is to be confronted with the 
value of the bulk binding energy per nucleon in nuclear matter of
$-15.75\,{\rm MeV}$ \cite{nuclear-matter}. 
Since the electric charge density is given by $e\,\rho(r)$, 
the electric mean square radius is given by
$\la r^2_{\rm ch}\ra=\int d^3r\,r^2\rho(r)$ 
(recall that $\int d^3r\,\rho(r)=1$). The model yields
$\la r_{\rm ch}^2\ra^{1/2} = 0.714\,{\rm fm}$, which underestimates 
the experimental value by $20\,\%$ but has the right order of magnitude. 
This model could be elaborated to give a more realistic description. 
However, the modest goal of Ref.~\cite{BialynickiBirula:1993ce} was to 
show that the model of the proton with the parameters (\ref{eq:parameters}) 
is ``not completely out of touch with reality.'' 
This is sufficient for our purposes.

\begin{figure}[b!]
  \begin{center}
    \includegraphics[width=0.25\textwidth]{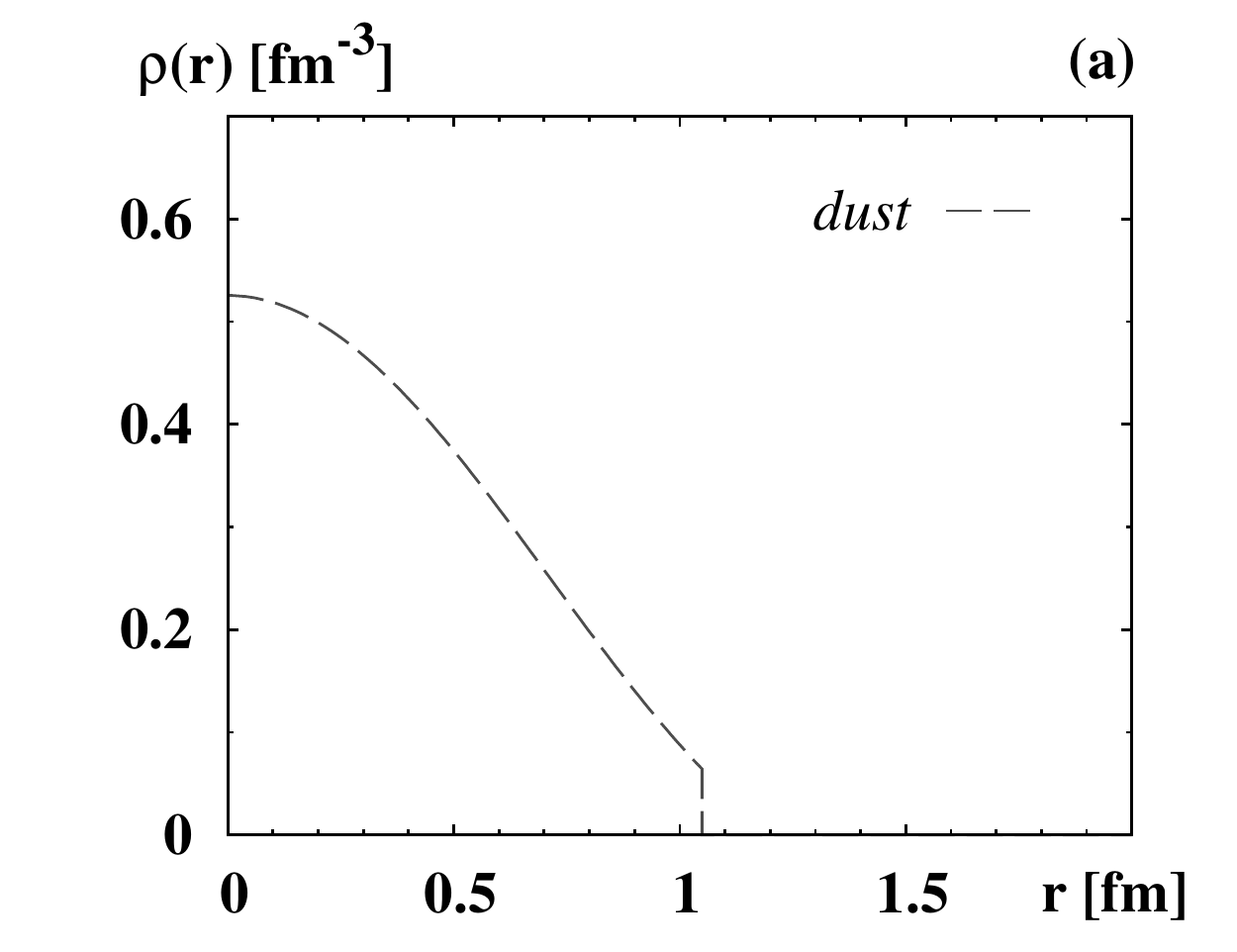}%
    \includegraphics[width=0.25\textwidth]{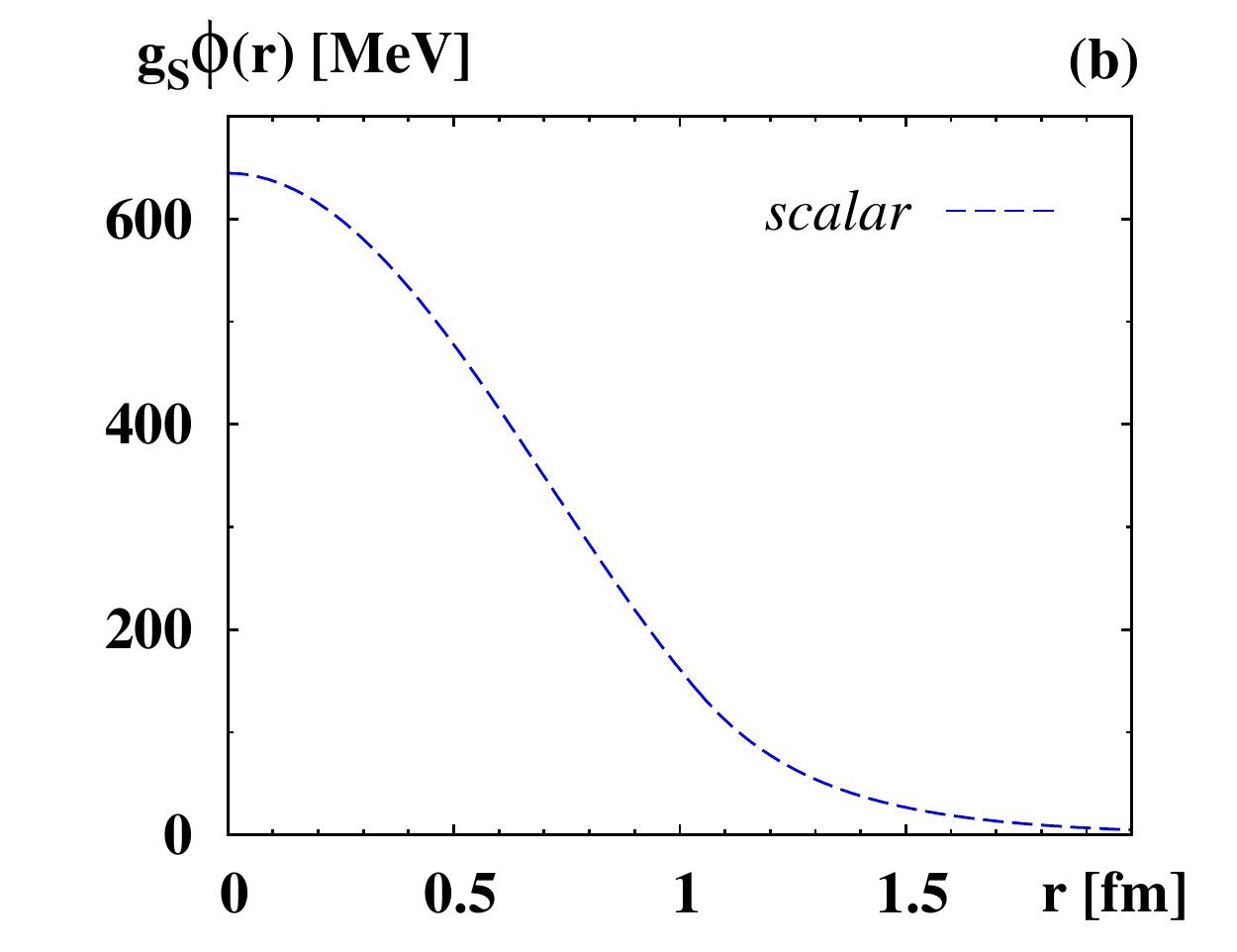}%
    \includegraphics[width=0.25\textwidth]{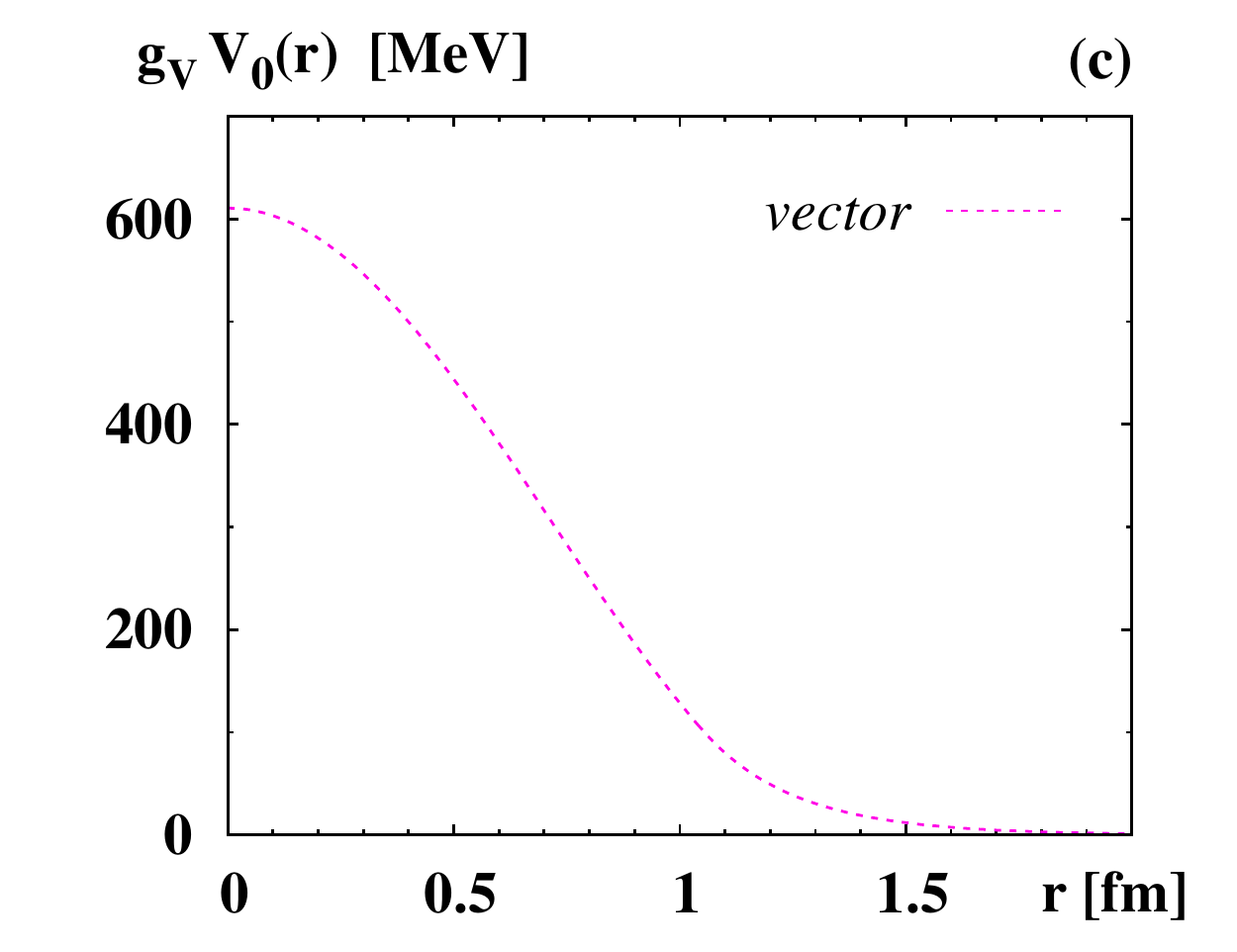}%
    \includegraphics[width=0.25\textwidth]{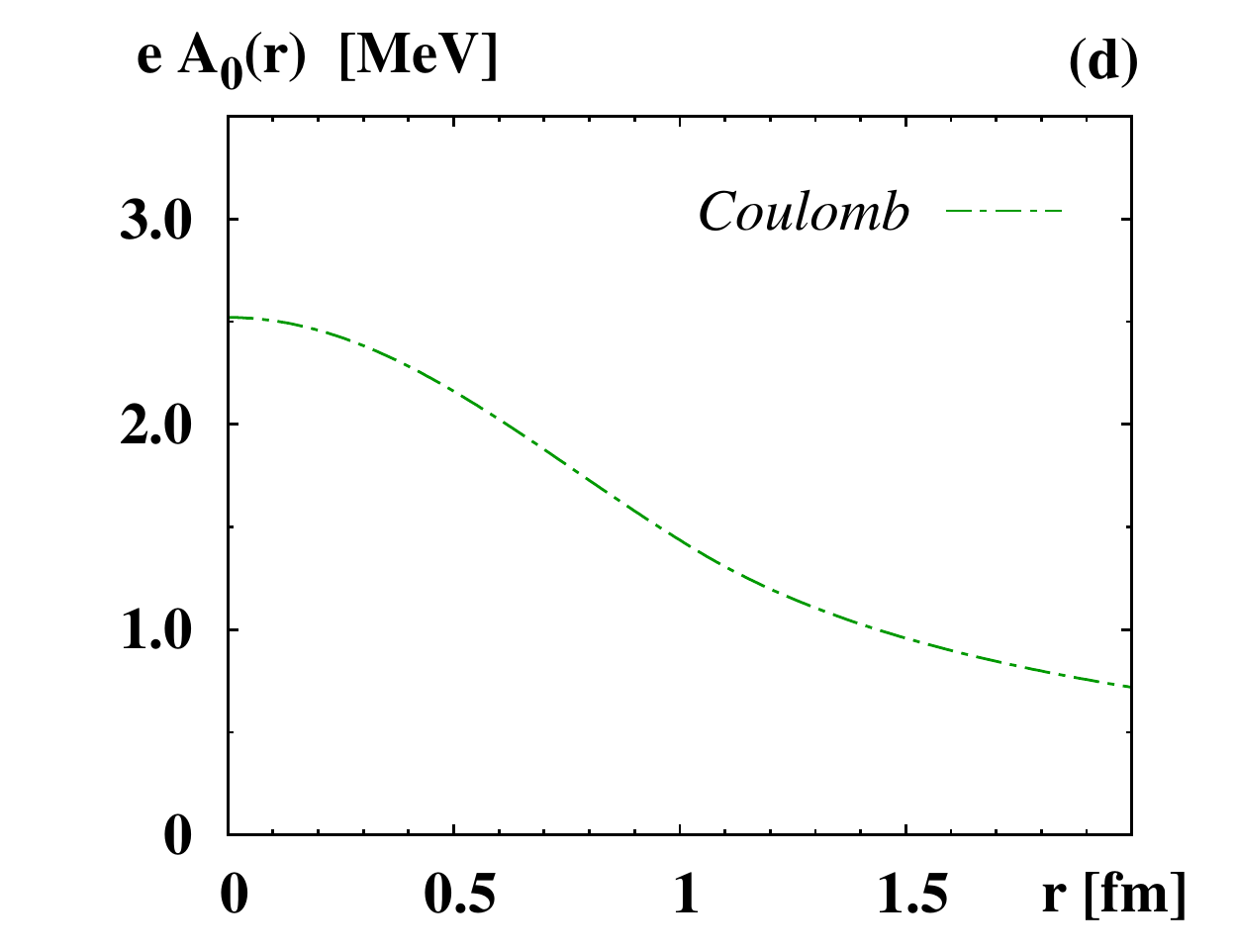}%
  \end{center}
  \caption{\label{fig1}
    Density of matter $\rho(r)$,
    scalar potential $g_{S}\phi(r)$,
    vector potential $g_V V_0(r)$, and
    Coulomb potential $eA_0(r)$ as functions of $r$.
    The density $\rho(r)$ drops to zero at $R=1.05\,{\rm fm}$.
    The potentials are multiplied by their respective coupling 
    constants such that the results in panels (b--d) are 
    potential energies given in units of MeV.}	
\end{figure}

In Fig.~\ref{fig1} we show the density $\rho (r)$, and the potentials 
$\phi(r)$, $V_0(r)$, $A_0(r)$ as functions of $r$. The jump in the 
matter distribution $\rho (r)$ shows that the dust particles 
are inside the radius $R=1.05\,{\rm fm}$ and there is no matter 
outside. The potentials in Figs.~\ref{fig1}b-d are scaled with 
their respective coupling constants such that they all have the 
same unit MeV and can be compared. The scalar potential 
$g_S\,\phi(r)$ and vector potential $g_VV_0(r)$ are associated 
with strong forces and are 2 orders of magnitude larger than 
the Coulomb potential $eA_0(r)$.
In the inner region, $g_S\phi(r)$ is somewhat larger than
$g_VV_0(r)$. At a larger $r$ it is the opposite: both potentials
decay at a rate of $\sim\,\exp(-m_i r)/r$ with $m_i=m_S$ and $m_V$ 
respectively, but $m_S<m_V$ so the more massive $V_0(r)$ 
has a shorter range compared to $\phi(r)$. 
The Coulomb potential $e A_0(r)$ is small in the interior 
region, but becomes the dominant field in the outer region 
thanks to its long range $A_0(r) \sim 1/r$. 
These results have already been discussed in 
Ref.~\cite{BialynickiBirula:1993ce}. 
In the next section we will discuss how the different fields contribute 
to the mass of the proton, and how the internal forces inside
the proton balance each other.

\section{The energy momentum tensor of the classical model}
\label{Sec-3:EMT-in-model}

The energy momentum tensor in the classical model of the
proton was derived in Ref.~\cite{BialynickiBirula:1993ce}
and is given by 
\begin{eqnarray}
         T^{ \mu \nu} & = &
         (m-g_S\,\phi)\rho u^\mu u^\nu +
         F^{\mu\rho}{F_\rho}^{\nu}  + \frac{1}{4}\,g^{ \mu \nu}\,F_{ \kappa\rho}F^{ \kappa\rho} + 
         \partial^\mu\phi\,\partial^\nu\phi - g^{\mu\nu}\,
         \biggl(\frac{1}{2}\partial_\rho\phi\,\partial^\rho\phi - \frac{1}{2}m_S^2\phi^2\biggr)  
         \nonumber\\
         &  + &
         G^{\mu\rho}{G_\rho}^{\nu} + m_V^2 V^\mu V^\nu + g^{\mu\nu}
         \biggl(\frac{1}{4}G_{\kappa\rho}G^{\kappa\rho}-\frac{1}{2}\,m_V^2 V_\rho V^\rho\biggr)   
         \label{eq:EMT}
\end{eqnarray}
This expression needs to be evaluated for the static solution where
$u^\alpha=(1,0,0,0)$,
$\phi=\phi(r)$,
$V^\alpha=(V_0(r),0,0,0)$,
$A^\alpha=(A_0(r),0,0,0)$.
In the following, we discuss the different components of the EMT, and 
focus initially on the region $r\le 2\,{\rm fm}$. The long-distance properties 
of the EMT at $r>2\,{\rm fm}$ will be discussed in the next section.

\subsection{Energy density \boldmath $T_{00}(r)$}

The $00$-component of the EMT describes the energy density.
Evaluating the $00$-component in Eq.~(\ref{eq:EMT}) yields 
\be\label{eq:T00}
	T_{00}(r) = ( m-g_{S} \phi  )  \rho 
        + \frac{1}{2} ( \vec{\nabla}  \phi  )^{2}
        + \frac{1}{2}m_{S}^{2} \phi ^{2}
        + \frac{1}{2} ( \vec{\nabla} V_0 ) ^{2}
        + \frac{1}{2}m_V^{2}V_0^{2}
        + \frac{1}{2} ( \vec{\nabla} A_{0} )^{2}  \,.
\ee
The mass of the solution is defined as $M=\int d^3r\,T_{00}(r)$.
Integrating the energy density (\ref{eq:T00}) over space, 
exploring the normalization $\int d^3r\,\rho(r)=1$, performing 
partial integrations and using 
Eqs.~(\ref{eq:minusdelmw}-\ref{eq:minusdelazero}), one obtains
\cite{BialynickiBirula:1993ce}
\ba
	M 
        &=&  m - \int d^3r\,g_{S}\phi\,\rho 
             + \frac{1}{2} \int  d^{3} r \biggl( 
             - \phi  \Delta  \phi +m_{S}^{2} \phi ^{2}
             - V_0 \Delta V_0+m_V^{2}V_0^{2}
             - A_{0} \Delta  A_{0} \biggr) \nonumber\\
        &=&  m + \frac{1}{2} \int  d^{3}r \biggl(
             -\,g_{S} \,\phi +g_V V_0 +e\,A_{0}\biggr)\rho \nonumber\\
        &=&  m + E_B \,. \label{eq:etot} \phantom{\int}
\ea
The potentials $g_{S} \phi(r)$, $g_V V_0(r)$, $eA_{0}(r)$ are 
positive, see Fig.~\ref{fig1}.
The intermediate step in (\ref{eq:etot}) shows that the 
scalar field makes a negative contribution to the total energy,
lowers the binding energy, and makes the system more strongly 
bound. In contrast to this, $g_V V_0(r)$ and $eA_{0}(r)$ enter with
positive signs, i.e.\ make the system less strongly bound.
These observations are not surprising and reflect the
well-known facts that scalar forces are attractive, and
vector forces (for equal sign charges) repulsive. 
We will come back to this point below 
when discussing the stress tensor. 

As mentioned in Sec.~\ref{Sec-2:model}, numerically, 
$E_B = -15.71\,{\rm MeV}$. This value can be compared to 
the bulk binding energy per nucleon in nuclear matter 
\cite{BialynickiBirula:1993ce}. We remark that alternatively
one could define the parameter $m$ in (\ref{eq:parameters}) 
to be a ``bare nucleon mass'' such that $m+E_B$ would be 
the physical mass of the free proton. 
Here we use the original version of the model as formulated 
in Ref.~\cite{BialynickiBirula:1993ce} and refrain from such 
a redefinition of model parameters.

\begin{figure}[b!]
  \begin{center}
    \includegraphics[width=0.25\textwidth]{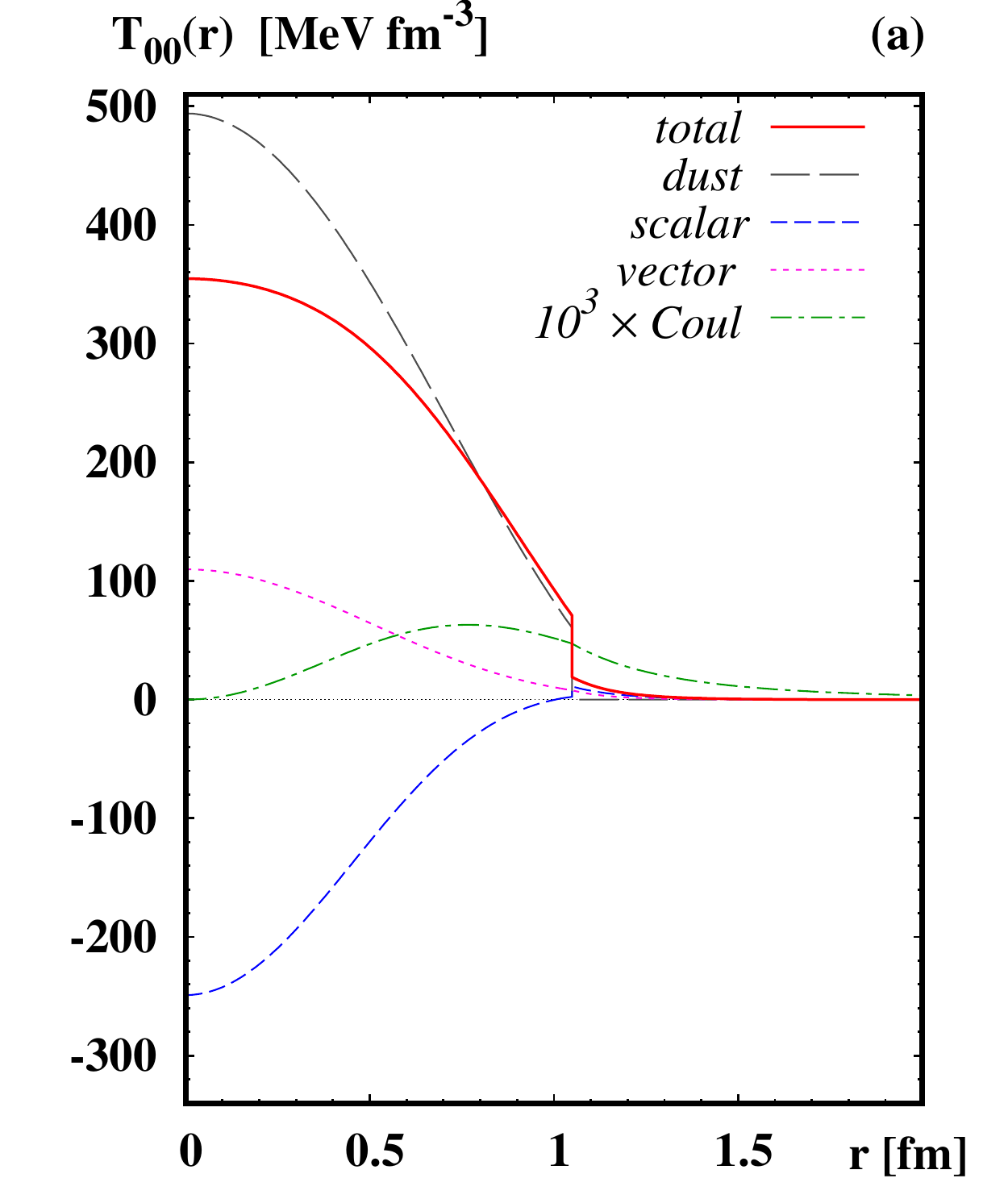}%
    \includegraphics[width=0.25\textwidth]{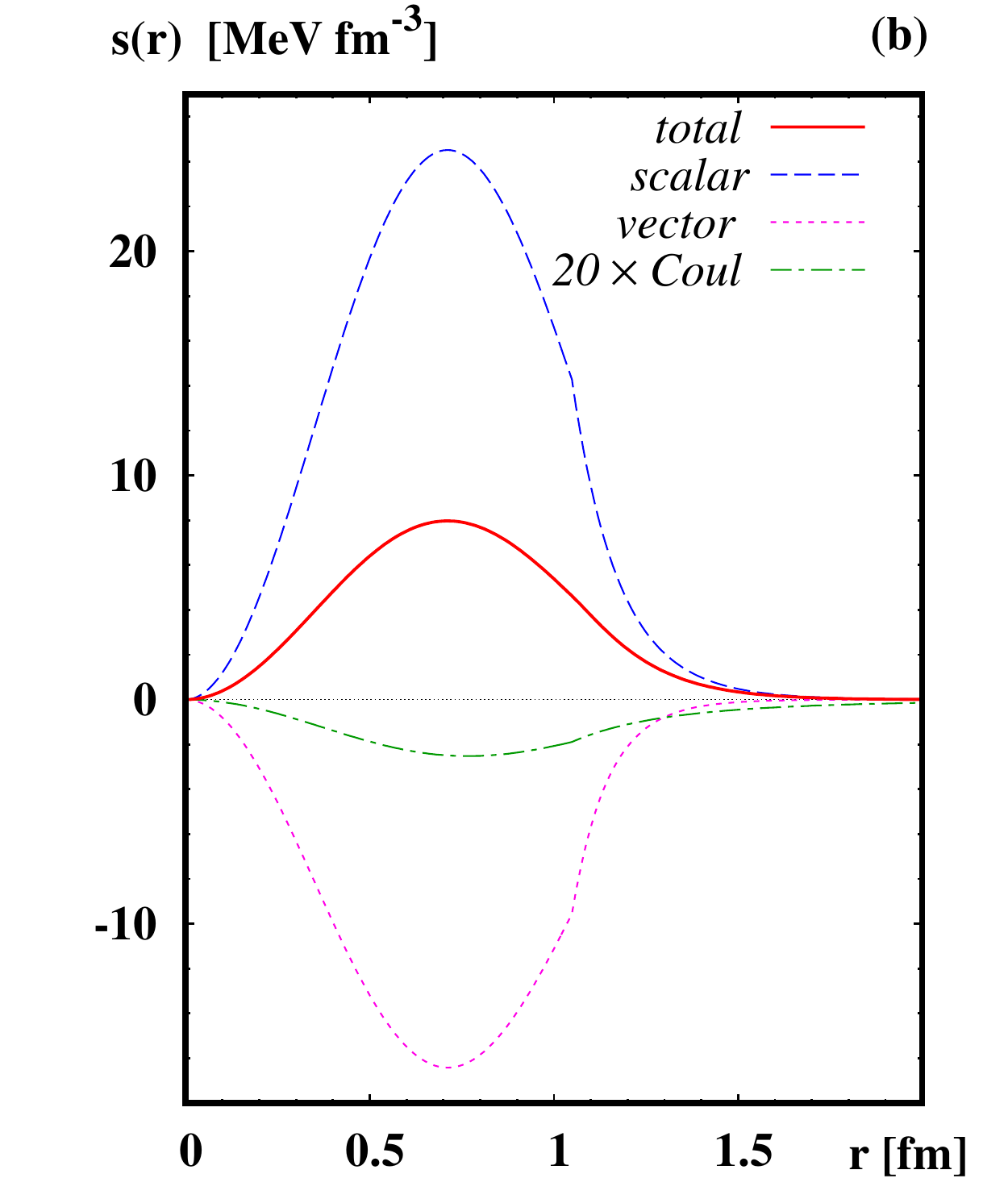}%
    \includegraphics[width=0.25\textwidth]{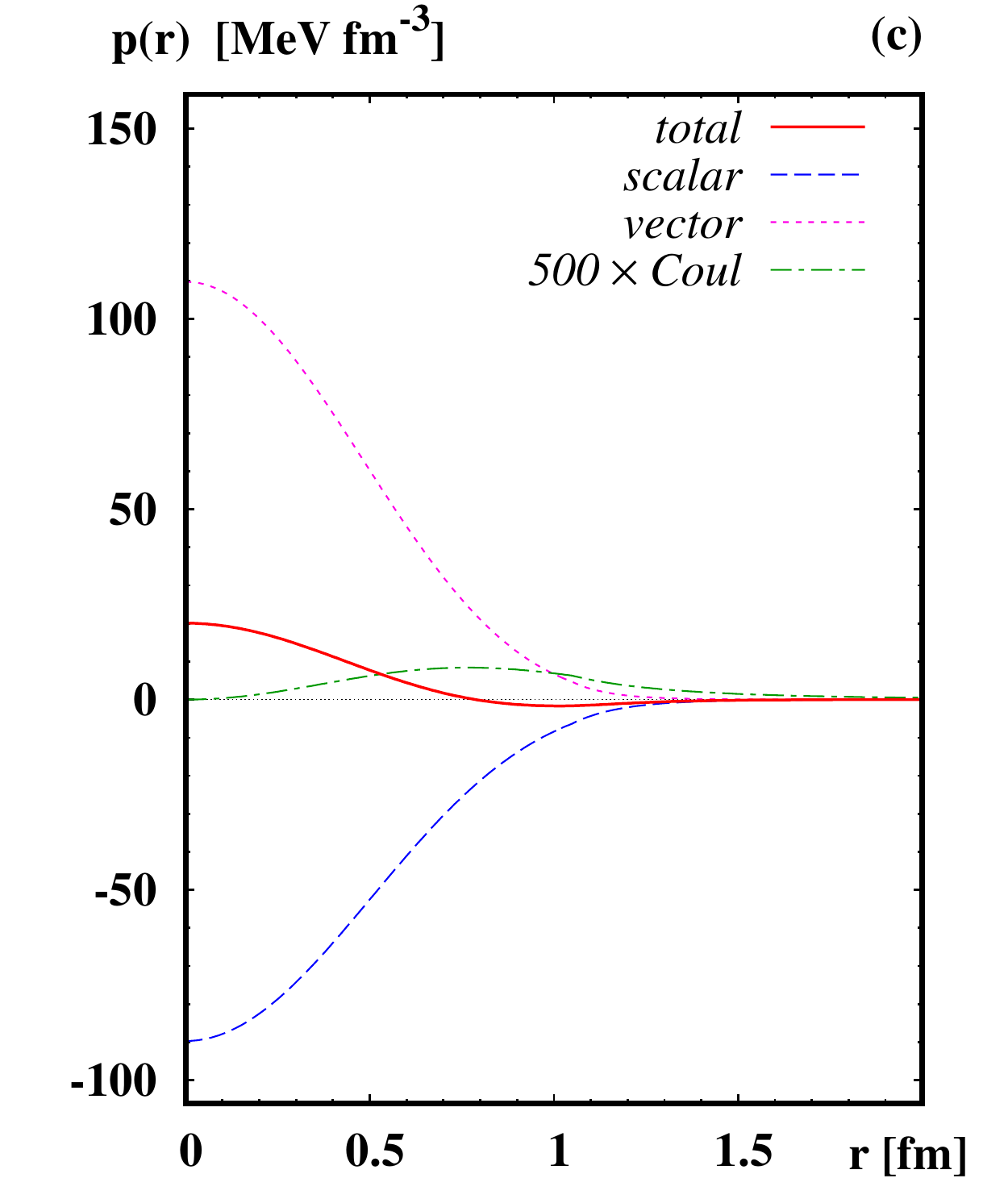}%
    \includegraphics[width=0.25\textwidth]{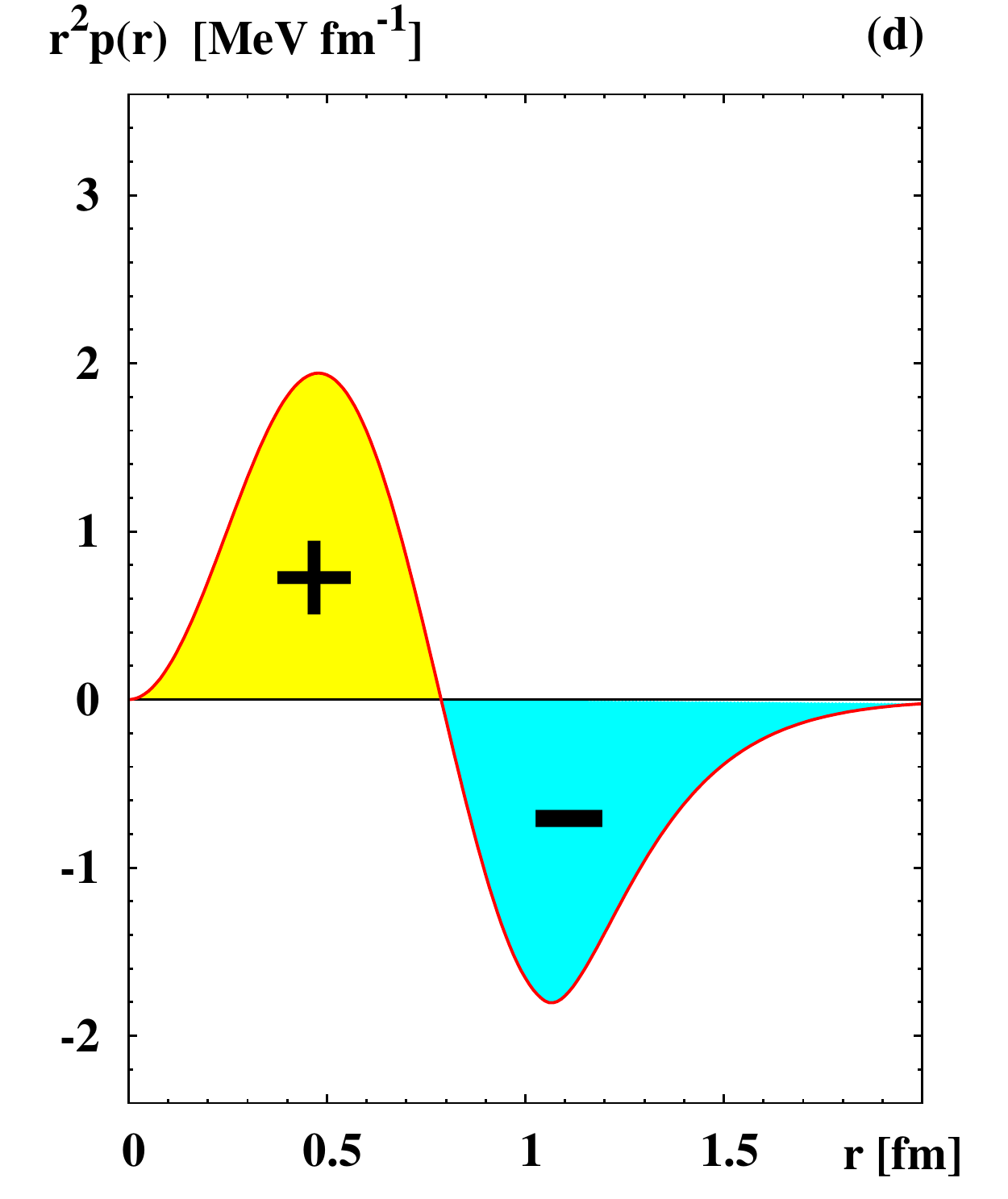}
  \end{center}
  \caption{\label{fig2}
    (a--c) EMT densities $T_{00}(r)$, $s(r)$, $p(r)$
    as functions of $r$ and the respective contributions
    from the dust particles, scalar, vector and
    Coulomb fields. $T_{00}(r)$ has a jump at $R=1.05\,{\rm fm}$, 
    while $s(r)$ and $p(r)$ exhibit kinks but remain continuous. 
    The very small Coulomb contribution is multiplied by the 
    indicated factors to make it visible on the scales of the plots. \\
    (d) Illustration of how the von Laue condition in Eq.~(\ref{eq:Laue}) 
    is satisfied, see text.}	
\end{figure}

It is instructive to discuss the individual contributions to 
the energy density which we define as follows
\begin{alignat}{2}
     T_{00}^{\rm dust}(r) & = m\,\rho(r)\,, \nonumber\\
     T_{00}^{\rm scal}(r) & = \frac12(\vec{\nabla}\phi)^2
                         + \frac12m_s^2\phi^2 -\,g_{S}\phi\,\rho\,\nonumber\\
     T_{00}^{\rm vect}(r) & = \frac12(\vec{\nabla}V_0)^2
                         + \frac12m_V^{2}V_0^{2} \,,\nonumber\\
     T_{00}^{\rm Coul}(r) & = \frac{1}{2} ( \vec{\nabla} A_{0} )^{2}\,, 
\end{alignat}
such that they add up to the total $T_{00}(r)$ in Eq.~(\ref{eq:T00}). 
Recalling that $\int d^3r\,\rho(r)=1$, we see that the dust particles
contribute $m$ and by far the most to $M$. The relatively small value 
$E_B = -15.71\,{\rm MeV}$ may give the incorrect impression that the 
contributions from the fields to $M$ are small. However, these 
contributions are given by
\be
    \int d^3r\,T_{00}^i(r)=\begin{cases} 
               -180.45\,{\rm MeV} & \quad {\rm for} \quad i = {\rm scalar}, \\
     \phantom{-}163.79\,{\rm MeV} & \quad {\rm for} \quad i = {\rm vector}, \\
     \phantom{-16}0.95\,{\rm MeV} & \quad {\rm for} \quad i = {\rm Coulomb}.  
     \end{cases}
\ee
Thus, the relatively small value of the binding energy is the 
result of large cancellations between different contributions. 
$T_{00}^{\rm scal}(r)$ is the only contribution which exhibits a 
discontinuity at $r=R$ and is negative in the inner region. 
The energy density and its individual contributions
are plotted in Fig.~\ref{fig2}a.

\subsection{The \boldmath $T_{0k}$ components}

For the $0k$-components of the EMT we obtain $T^{0k}=0$ which
is not surprising. 
This is because we deal with a static solution. Since there 
is no rotation in the system, the classical angular momentum
$J^i=\int d^3r\,\epsilon^{ijk}x^j T^{0k}$ of
the system is zero. At this point, we disregard the fact that
the proton has spin $\frac12$. In principle, if interested,
one could treat our classical solution as a soliton and use 
standard quantization techniques to assign a definite spin
\cite{Rajamaran}. 

\subsection{The stress tensor \boldmath $T_{ij}$}

Finally, for the $ij$-components of the EMT we obtain the result
\begin{eqnarray}
	T^{ij}(r) 
	&=& e_{r}^{i}e_{r}^{j}
	\biggl(\phi'(r)^{2} - A_{0}'(r)^{2} - V_0^{'}( r)^{2} \biggr)
	\nonumber\\
 	&+&\;\frac{\delta^{ij}}{2}\biggl(
	(\vec{\nabla} A^{0})^{2} 
	- \frac{1}{2}(\vec{\nabla}\phi)^{2} 
	- \frac{1}{2}m_{S}^{2}\phi^{2} 
	+ \frac{1}{2}(\vec{\nabla} V_0)^{2} 
	+ \frac{1}{2}m_V^{2}V_0^{2}\biggr)
\end{eqnarray}
In general, the stress tensor can be expressed in terms of a 
traceless part associated with shear forces $s(r)$ and a trace 
associated with the pressure $p(r)$ as follows
\be
       T^{ij}= \biggl( e_{r}^{i}e_{r}^{j}-\frac{1}{3}\delta ^{ij}\biggr)s(r)
       + p(r)  \delta ^{ij}\,,
\ee
where $e_r^i$ is the unit vector in the radial direction.
The model results for $s(r)$ and $p(r)$ are given by
\ba
    s(r) &=& \phi '(r)^{2}-V_0'(r)^{2}-A_{0}'(r)^{2} \,,\label{eq:s} \\
    p(r) &=& \frac{1}{6}A_{0}'(r)^{2}-\frac{1}{6} \phi '(r)^{2}
    -\frac{1}{2}m_{S}^{2} \phi ^{2}+\frac{1}{6}V_0'(r)^{2}
    +\frac{1}{2}m_V^{2}V_0^{2} \,.
    \label{eq:p}
\ea
Due to the EMT conservation, $\partial _\mu T^{\mu\nu}=0$, 
the shear forces and pressure are not independent of each other but 
connected by the differential equation \cite{Polyakov:2018zvc}
(we leave here the space dimension $n=3$ general for later purposes)
\be\label{eq:thedifferentialequation}
\frac{n-1}{r}s(r)+\frac{n-1}{n}s^\prime(r)+p^\prime(r)=0\,.
\ee
 Another consequence of the EMT conservation 
is the von Laue condition 
\be
	 \int _{0}^{\infty}dr~r^{2}p(r)=0 \,.
         \label{eq:Laue}
\ee
This is a necessary (but not sufficient) condition for stability,
and requires that internal forces inside a system must exactly 
balance each other. If the integral in \eqref{eq:Laue} was
positive (negative), then the system would explode (implode).

Notice that the dust particles do not contribute to the stress
tensor densities $s(r)$ and $p(r)$. This is naturally explained
in the hydrodynamic interpretation of the model, where the dust 
particles can be viewed as an ideal pressureless fluid of density 
$\rho(r)$, which flows (without dissipation) with the 4-velocity 
$u^\mu$ \cite{BialynickiBirula:1993ce}. Therefore, the densities 
$s(r)$ and $p(r)$ receive contributions only from the fields.

The scalar field makes the largest contribution to $s(r)$, 
see Fig.~\ref{fig2}b, which is positive.
The contributions from vector and Coulomb fields are both
negative. Not surprisingly, the Coulomb field contribution is 
rather small. The shear forces behave like $s(r)\propto r^2$ 
at small $r<0.1\,{\rm fm}$, and exhibit a global maximum at
$r=0.711\,{\rm fm}$, which is numerically close (within $0.1\,\%$)
but not the same value as the charge radius 
$\la r_{\rm ch}^2\ra^{1/2}=0.714\,{\rm fm}$. 
For a large nucleus in the liquid drop model, $s(r)$ would be a 
$\delta$-function centered at the edge of the nucleus (with the 
coefficient in front of the $\delta$-function given by the surface 
tension) \cite{Polyakov:2018zvc}. The result for $s(r)$ remotely 
resembles a strongly smeared out $\delta$-function.
This reflects the fact that the proton has no sharp edge, and is 
a much more diffuse object than a nucleus.

The pressure is positive in the inner region, changes sign at
$r=0.788\,{\rm fm}$ and is negative thereafter, see Fig.~\ref{fig2}c. 
The sign convention is such that $p(r)>0$ means repulsive forces are
directed towards the outside, while $p(r)<0$ means attractive forces are
directed towards the inside.
The shape of the total pressure distribution is largely due to 
the cancellation between the large contributions from scalar
and vector fields. In the inner region, the repulsive vector forces 
are stronger than the attractive scalar forces. In the outer region, 
it is vice versa since, due to $m_V>m_S$, the range of the vector 
forces is shorter.
Throughout the region plotted in Fig.~\ref{fig2}c, the Coulomb contribution 
plays a minor role (but is not negligible, see below) and contributes
to $p(r)$ with the same sign as the vector field.

In order to attest the consistency of our calculation, we notice 
that inserting the expressions (\ref{eq:s},~\ref{eq:p}) for $s(r)$ 
and $p(r)$ into Eq.~(\ref{eq:thedifferentialequation}) yields 
$\frac{2}{3}s^\prime(r) +\frac{2}{r}s (r) +p^\prime(r)=
\vec{e}_r[\rho\,\vec{F}]=0$ due to Eq.~(\ref{eq:force-equilibrium}).
Also, the von Laue condition (\ref{eq:Laue}) holds which was 
proven in \cite{BialynickiBirula:1993ce}, and is illustrated 
in Fig.~\ref{fig2}c. If we define  the individual contributions 
to the pressure as
\ba
  p_{\rm scal}(r) &=& -\frac16\,\phi'(r)^2-\frac12\,m_S^2 \phi(r)^2
  \,,\nonumber\\
  p_{\rm vect}(r) &=& \phantom{-}\frac16\,V_0'(r)^2+\frac12\,m_V^2V_0(r)^2 
  \,,\nonumber\\
  p_{\rm Coul}(r) &=& \phantom{-}\frac{1}{6}\,A_{0}'(r)^{2} \,,
\ea
then the contributions from the scalar, vector and Coulomb field to the
von Laue integral in Eq.~(\ref{eq:Laue}) are 
\be
       \int dr\;r^2p_i(r) = \begin{cases} 
               -10.916\,{\rm MeV} & \quad {\rm for} \quad i = {\rm scalar}, \\
     \phantom{-}10.891\,{\rm MeV} & \quad {\rm for} \quad i = {\rm vector}, \\
     \phantom{-1}0.025\,{\rm MeV} & \quad {\rm for} \quad i = {\rm Coulomb}.  
     \end{cases}       
     \label{eq:Laue-in-detail}
\ee
These results mean that with scalar forces alone, the system would
implode, while with vector (or Coulomb) forces alone it would explode.
Clearly, in Eq.~(\ref{eq:Laue-in-detail}) the contribution of the Coulomb 
force is minuscule compared to that of the strong forces, but {\it not}
negligible, for this system would implode without the Coulomb force. 

Some comments regarding the size of the forces are in order. In our model, 
the pressure in the center of the proton is about $20\,\rm MeV/fm^3$. This 
is about an order of magnitude less than in the chiral quark soliton model 
\cite{Goeke:2007fp}. This result is expected and plausible for the following 
reason. The forces in the model of Ref.~\cite{Goeke:2007fp} are the strong 
forces acting between quarks. In contrast to this here, the strong forces 
(the massive scalar and vector fields) are modeled using nuclear physics 
phenomenology. Such ``residual nuclear forces'' are about an order of
magnitude weaker than the strong forces among quarks inside the proton, 
and this is what we observe. 

To make an intermediate summary, the description of the EMT in the 
classical proton model is internally consistent since the relations
(\ref{eq:thedifferentialequation},~\ref{eq:Laue}) hold. The size of
the internal forces is what one would expect from a model with forces
which have the strength of residual nuclear forces.
The results discussed so far reflect the same features as 
encountered  in other EMT studies in strongly interacting systems, 
including the sign patterns for the EMT densities in  
Fig.~\ref{fig2}. 

\newpage
\section{EMT densities and long-range forces}
\label{Sec-4:EMT+long-range-forces}

The particular feature of the model used in this work is that it
explicitly includes long-range (Coulomb) forces. Before we study
this aspect in detail, let us briefly review several common features
observed in prior EMT studies of strongly interacting systems governed
by short-range forces \cite{Polyakov:2018zvc}. For ground state solutions,
in systems governed by strong short-range forces, 
the following common features were observed so far:
\begin{enumerate}
\item the shear forces $s(r)$ are positive at all $r$,
\item the pressure $p(r)$ exhibits one node at $r_0$ with
$p(r)>0$ for $r<r_0$ and $p(r)<0$ for $r>r_0$,
\item the combination $\frac23\,s(r)+p(r)$, which is referred 
to as normal force (per unit area), is always positive, i.e.\
\be
      \frac23\,s(r)+p(r)>0\,. \label{eq:positivity-norm-force}
\ee
\end{enumerate}
Some comments are in order. We are not aware of a rigorous proof 
of the property (i), though it is plausible given the connection 
of $s(r)$ to surface tension and surface energy, which are positive 
in stable hydrostatic systems \cite{Polyakov:2018zvc}.
The positivity of $s(r)$ was observed in all studies so far.
The property (ii) arises because $p(r)$ must have at least 
one node to comply with the von Laue condition (\ref{eq:Laue}),
and ground states exhibit a single node. The pattern in
Fig.~\ref{fig2}d follows from mechanical stability arguments:
repulsive forces are required in the inner region to prevent 
collapse and attractive forces in the outer region to bind 
the system  \cite{Polyakov:2018zvc}.
For excited states, the pressure can exhibit several nodes, but the
pattern with $p(r)>0$ in the center and $p(r)<0$ at large distances
remains \cite{Mai:2012cx}. 
The property (iii) is a mechanical stability criterion
and means that the radial forces $T^{ij}\,d A^j_r$, where 
$d A^j_r=e_r^j \,r^2d\Omega$, are directed towards the outside,
and the point where they vanish (if we deal with a finite size system)
marks the ``edge'' of the system \cite{Polyakov:2018zvc,Neubelt:2019sou}.

As long as we consider distances $r\lesssim 2\,{\rm fm}$,
the EMT densities in the classical proton model exhibit the 
properties (i--iii) as observed in prior studies. But the
situation changes when we consider distances $r\gtrsim 2\,{\rm fm}$.

\subsection{Long-distance behavior of the EMT densities}

\begin{figure}[t!]
  \begin{center}
    \includegraphics[width=0.25\textwidth]{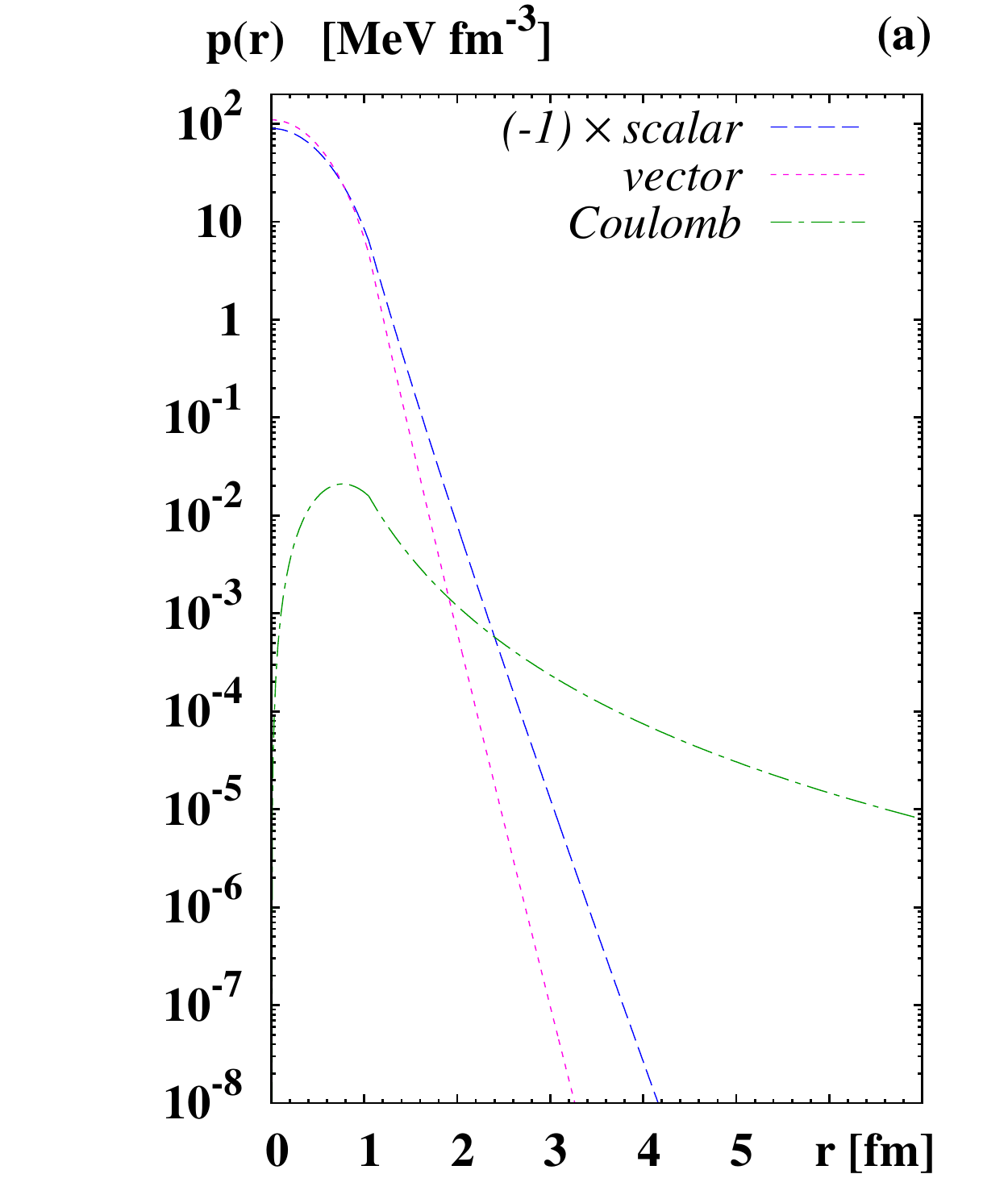}%
    \includegraphics[width=0.25\textwidth]{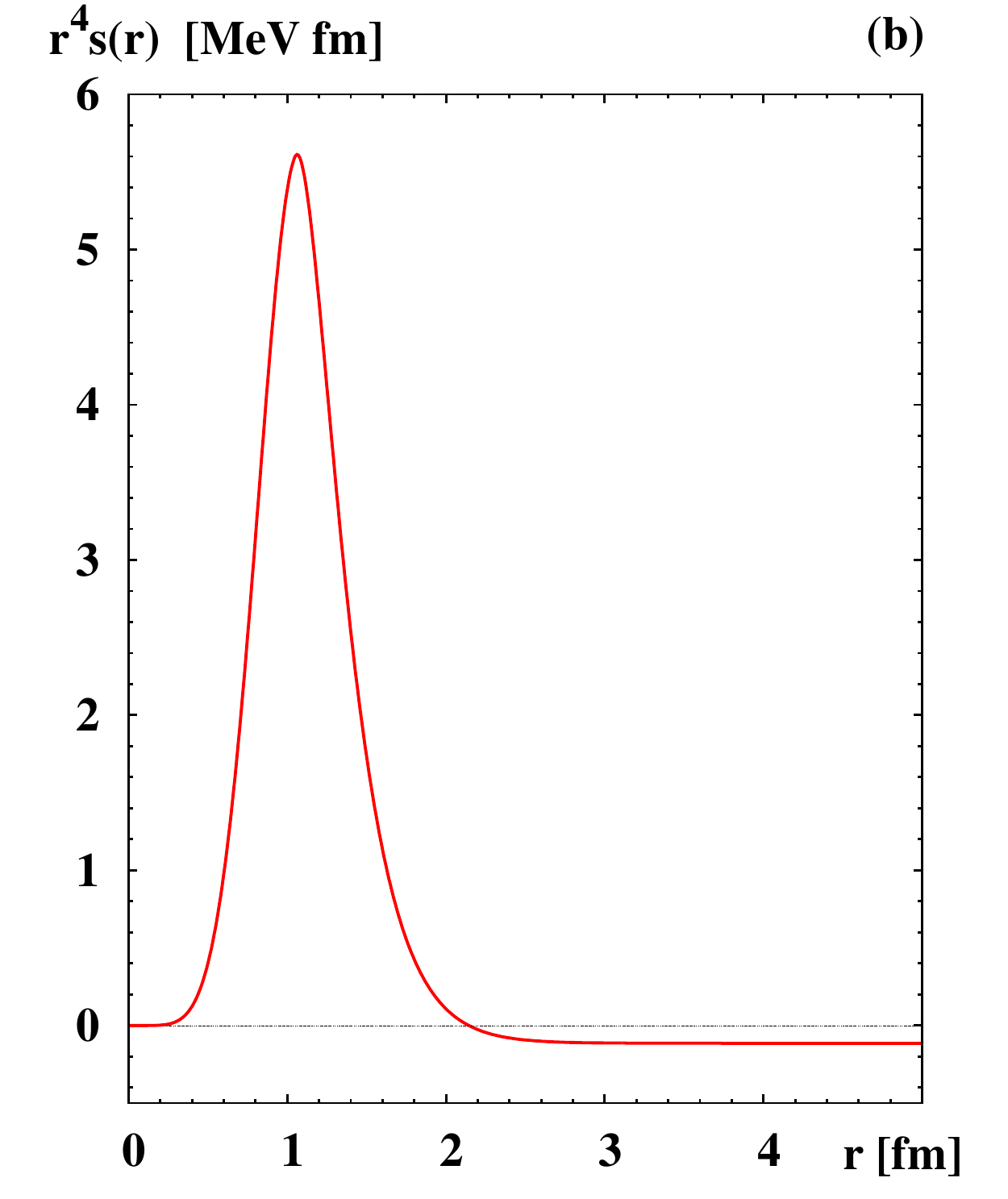}%
    \includegraphics[width=0.25\textwidth]{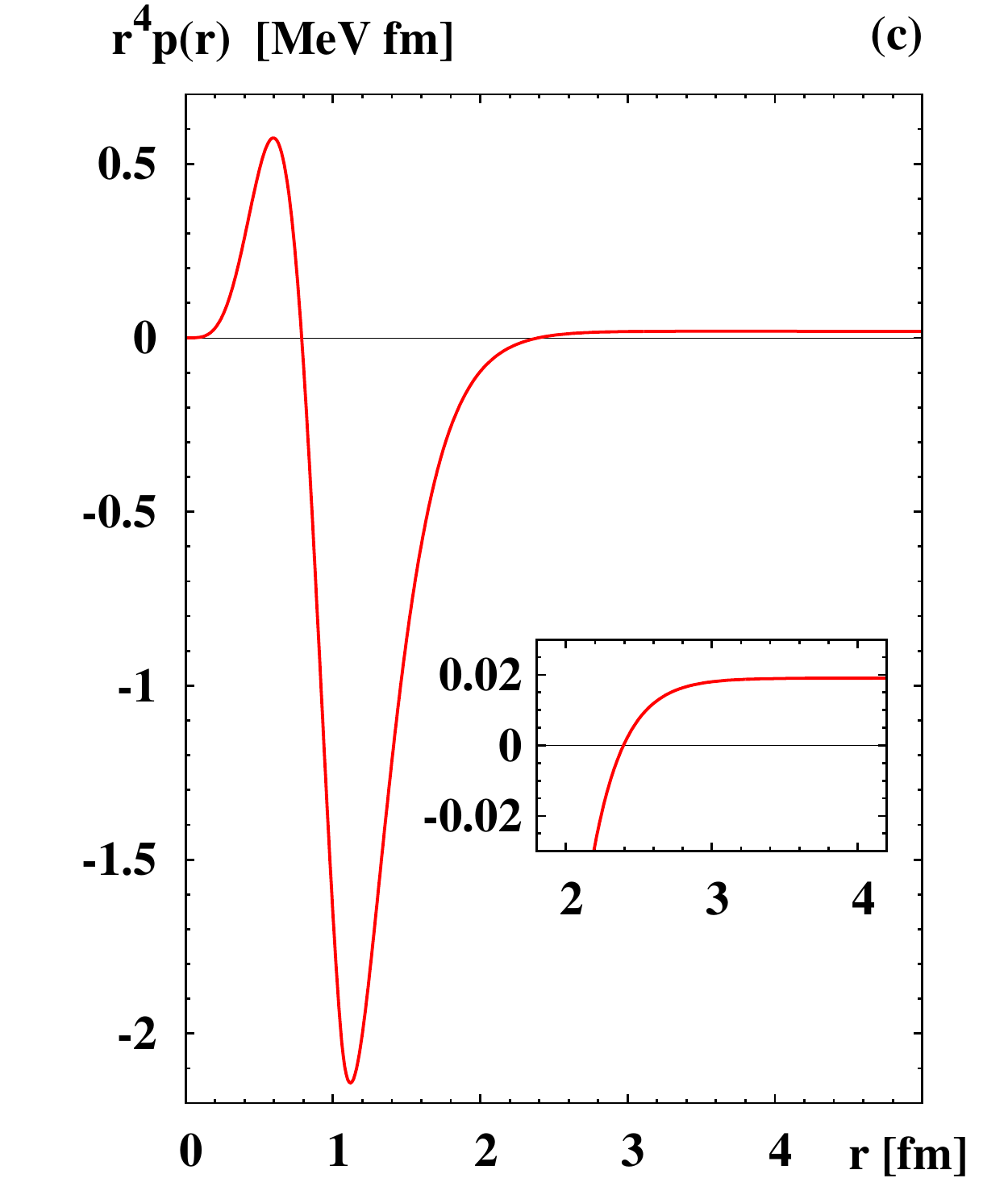}%
    \includegraphics[width=0.25\textwidth]{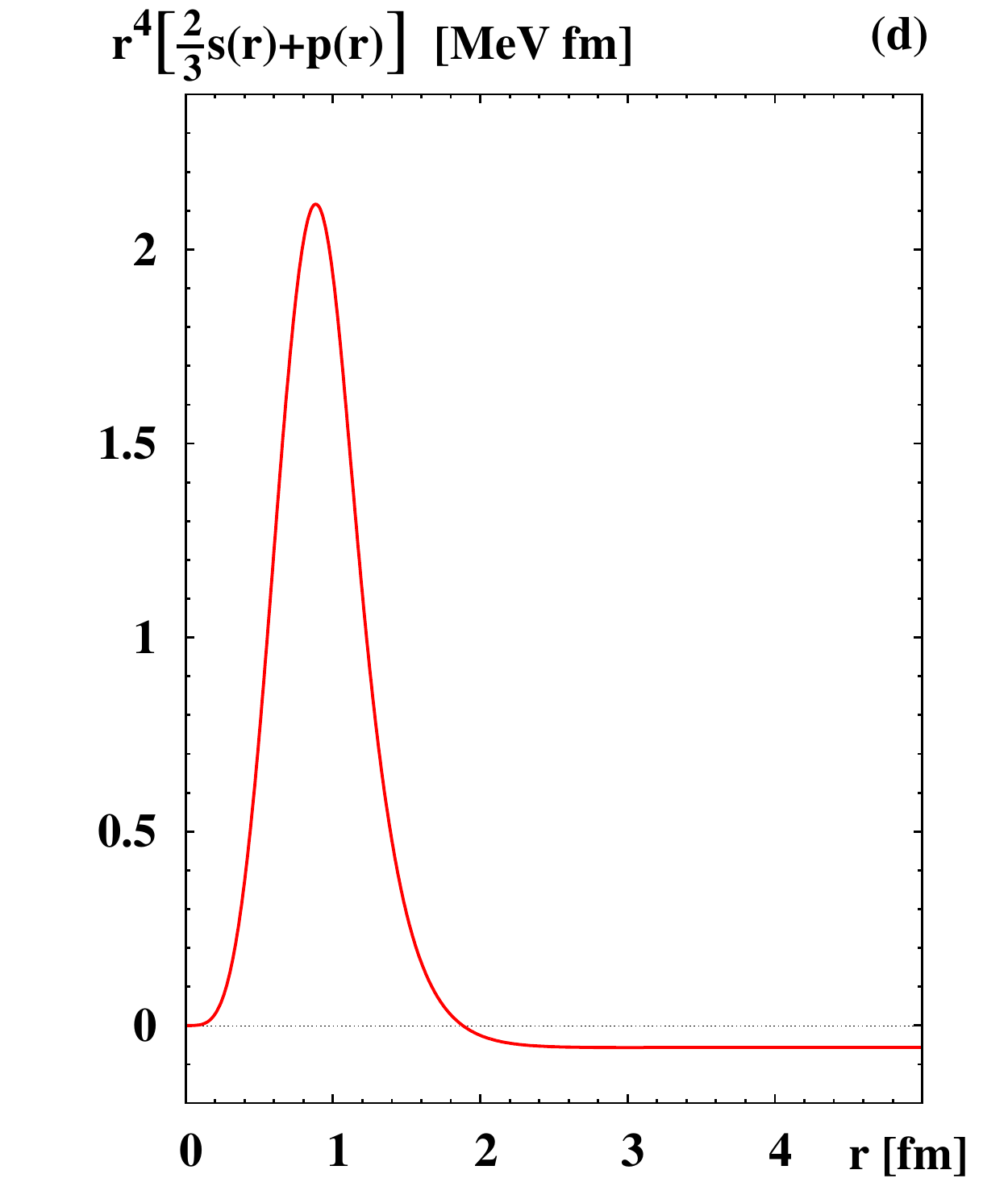}
  \end{center}
  \caption{\label{fig3}
    (a) Contributions to $p(r)$ from the fields on 
    a log-scale to visualize the dominance of the Coulomb field at 
    long distances. The total results for 
    (b) shear forces $s(r)$,
    (c) pressure $p(r)$, and
    (d) normal forces $\frac23sr(r)+p(r)$ multiplied by $r^4$
    such that the Coulomb contributions $\propto 1/r^4$ show as
    straight lines at long distances.}
\end{figure}

The dust distribution is confined to the region $r<R=1.05\,{\rm fm}$ 
and anyway does not contribute to $s(r)$ and $p(r)$. 
The behavior of the EMT densities at long distances is therefore
determined by the fields. The contributions of the fields representing
the strong forces decay exponentially at large distances
$r\gtrsim 2\,{\rm fm}$. Despite being very small in the inner region, 
the Coulomb contribution becomes comparable to the contributions of the 
strong fields at $r\sim\,$2--3$\,$fm. The Coulomb contribution is the dominating 
field at long-distances $r\gtrsim3\,{\rm fm}$ due to the slow, power-like 
$\frac1r$-decay of the Coulomb potential. 
This is illustrated in Fig.~\ref{fig3}a for the pressure; 
the situation is very similar for $T_{00}(r)$ and $s(r)$. 
Using the fine-structure constant in (\ref{eq:parameters}) we read 
off from Eqs.~(\ref{eq:T00},~\ref{eq:s},~\ref{eq:p}) the 
long-distance behavior of the densities
\ba 
	T_{00}(r)   &=& \frac12\;
	\frac{\alpha}{4\pi}\;\frac{\hbar c}{r^4} + \dots\,\nonumber\\
    s(r)        &=& -\,
    \frac{\alpha}{4\pi}\;\frac{\hbar c}{r^4} + \dots\,\nonumber\\
    p(r)        &=& \frac{1}{6}\;
    \frac{\alpha}{4\pi}\;\frac{\hbar c}{r^4} + \dots\,
    \label{eq:long-distance}
\ea
where the dots indicate subleading, exponentially suppressed
contributions from the strong interaction fields. 

The long-distance behavior (\ref{eq:long-distance}) of $T_{00}(r)$
does not show anything unusual: $T_{00}(r)>0$ for all 
$0\le r<\infty$ which was also observed in all prior studies. 
It is noteworthy that the $1/r^4$-decay of $T_{00}(r)$ at 
large $r$ guarantees the convergence of the total energy 
$M=\int d^3r\,T_{00}(r)$. But the mean square radius of the 
energy density $\la r_E^2\ra=\int d^3r\,r^2T_{00}(r)/M$ 
diverges and cannot be defined in this model. 

New features emerge for the stress tensor densities $s(r)$ 
and $p(r)$. From Eq.~(\ref{eq:long-distance}) we see that 
$s(r)$ is negative at large $r$. The classical proton model 
is in agreement with the property (i) and exhibits a positive
$s(r)$, see Fig.~\ref{fig2}b, throughout the region $0<r<r_{0,s}$.
But $s(r)$ changes sign at the point $r_{0,s}=2.144\,{\rm fm}$, 
and remains negative for $r>r_{0,s}$. 
The asymptotic expression for $s(r)$ in Eq.~(\ref{eq:long-distance}) 
works with an accuracy of $2\,\%$ or better for distances 
$r\gtrsim3.1\,{\rm fm}$.

Another new feature is that $p(r)$ is positive at large $r$, 
see Eq.~(\ref{eq:long-distance}). Throughout the region $0<r<r_{0,p}$,
the model conforms to the property (ii) and $p(r)$ exhibits 
the characteristic pattern: positive $p(r)$ in the inner region, 
a single node, and negative pressure in the outer region, 
see Figs.~\ref{fig2}c and \ref{fig2}d. 
But then at $r_{0,p}=2.394\,{\rm fm}$, the
pressure exhibits an {\it additional} (second) change of sign
after which it remains positive. The asymptotic expression 
(\ref{eq:long-distance}) for $p(r)$ works with an accuracy of 
$1\,\%$ or better beyond $r\gtrsim3.4\,{\rm fm}$.
We stress that the classical model describes a ground state
(and, in fact, no excited solutions exist in this model)
\cite{BialynickiBirula:1993ce}. But nevertheless $p(r)$ 
exhibits {\it two} nodes. 

Finally, the normal force $\frac23\,s(r)+p(r)$ is positive for 
$0\le r < r_{0,n}$ in agreement with condition (iii), until exhibiting 
a node at $r_{0,n}=1.881\,{\rm fm}$, after which it is negative, another 
new feature. Notice that $\frac23\,s(r)+p(r)\propto \frac{1}{r^4}+\dots$,
implying that the mechanical mean square radius 
$\la r^2_{\rm mech}\ra=\int d^3r\,r^2(\frac23s(r)+p(r))/\int d^3r\,(\frac23s(r)+p(r))$ diverges.

The 3 new features consist of
(i) a node in $s(r)$,
(ii) a second node in $p(r)$, and
(iii) a node in the normal force.
After the appearance of the nodes, the respective densities exhibit 
opposite signs as compared to prior studies. It is worth remarking 
that in Eq.~(\ref{eq:long-distance}), the asymptotics of $T_{00}(r)$ 
and $p(r)$ are such that $T^{00}(r)-3\,p(r)=0$, which reflects the
tracelessness of the EMT tensor ${T_\mu}^\mu = 0$ (in classical
electrodynamics). The asymptotic expressions (\ref{eq:long-distance}) 
for $s(r)$ and $p(r)$ satisfy the differential equation 
(\ref{eq:thedifferentialequation}) which is dictated by 
the conservation of the EMT.

The size of $s(r)$ and $p(r)$ is very small in the regions
where the new features occur. For instance, the second node
of $p(r)$ at $r_{0,p}=2.394\,{\rm fm}$ is beyond the range of 
Figs.~\ref{fig2}c and \ref{fig2}d. However, had we tried to 
show it there, then the second node would be hardly visible 
on the scales of the Figs.~\ref{fig2}c and \ref{fig2}d.
In order to visualize the new features, we multiply the 
respective densities by $r^4$ such that the Coulomb 
contributions proportional to $1/r^4$ appear as constant lines
at large $r$, see Figs.~\ref{fig3}b--\ref{fig3}d. 
Despite the factor $r^4$ which enhances the densities at
large $r$, the Coulomb contribution is small even in these
plots. In particular, it is so small in the case of $r^4p(r)$ 
that an insert is necessary (with the scale on the y-axis 
enlarged by a factor of 10) to clearly show the second zero of 
$p(r)$ in Fig.~\ref{fig3}c.

\subsection{The divergence of the \boldmath $D$-term}

The presence of the long-range electromagnetic forces also affects
 the $D$-term, which is an important particle property and on
the same footing as the mass, spin or electric charge 
\cite{Polyakov:2018zvc}.
The $D$-term has two equivalent definitions 
\ba
    D_s &=& 
              - \frac{2(n-1)}{n(n+2)}\,
              M \int d^n r\; r^2 s(r)\,,\label{eq:Ds}\\
    D_p &=&   M \int d^n r\; r^2 p(r) \,, \label{eq:Dp}
\ea
in terms of shear force and pressure where $n=3$ is the
space dimension, which we leave here general for later
purposes.
These expressions are equivalent due to the EMT conservation,
i.e.\ $D=D_s=D_p$ gives the same result. This can be proven
exploring the differential relation
(\ref{eq:thedifferentialequation}), cf.\ Ref.~\cite{Goeke:2007fp}.

In the classical model of the proton, the $D$-term is
undefined because the integrals in Eqs.~(\ref{eq:Ds},~\ref{eq:Dp})
diverge linearly due to the asymptotic behavior 
(\ref{eq:long-distance}) of $p(r)$ and $s(r)$ at large distances. 
Notice that the Figs.~\ref{fig3}b and \ref{fig3}c 
basically show the integrands in Eqs.~(\ref{eq:Ds},~\ref{eq:Dp}).
Even though the Coulomb contribution is very small, it clearly spoils 
the convergence of the integrals in Eqs.~(\ref{eq:Ds}) and (\ref{eq:Dp}).

This is a new feature not encountered in previous studies.
Typically in strongly interacting systems, the EMT densities
decay at long distances fast enough such that the integrals
defining the $D$-term converge. In quantum field theoretical models 
of the nucleon, the EMT densities exhibit an exponential fall-off at 
large $r$ for finite pion masses. In the chiral limit, 
when the Goldstone boson (pion) becomes strictly massless,
$s(r)$ and $p(r)$ behave like $1/r^6$ at large $r$, which is still
sufficient to guarantee a finite, well-defined $D$-term  \cite{Goeke:2007fp}.

In the remainder of this work we will address the following questions. 
Does the model constitute a mechanically stable solution?
Is it possible to obtain a prediction for the $D$-term in this model?
And, are our observations model-dependent or of general character?

\subsection{Mechanical stability in the model}

Even though constructed as a consistent classical mechanical model 
of the proton \cite{BialynickiBirula:1993ce}, we find that the mechanical 
stability criterion (\ref{eq:positivity-norm-force}) is not satisfied at 
large $r$. This issue needs to be resolved.
Let us stress that positivity of the normal force
(\ref{eq:positivity-norm-force}) is a stability criterion 
of mechanical continuum systems \cite{LLv7}. Care may be needed 
when carrying over such criteria to quantum systems \cite{Polyakov:2018zvc}. 
But here we deal with a classical continuum system and the 
criterion (\ref{eq:positivity-norm-force}) must hold. 

In this context, it is interesting to recall how the criterion
(\ref{eq:positivity-norm-force}) is used to determine the 
radius of a neutron star: 
density and radial pressure in the neutron star interior 
are governed by the Tolman-Oppenheimer-Volkoff equations which 
include general relativity effects, and are connected to each 
other by an equation of state of nuclear forces.  
The equation of state contains
information on the compressibility of nuclear matter. 
The solution of these equations yields the normal force
(which in  neutron star literature is often referred 
to as ``radial pressure'' or simply ``pressure'', not to be
confused with the pressure $p(r)$ in this work.)
The normal force is positive, but at some point it turns negative. 
This point marks the radius of the neutron star. Could we apply 
the same procedure to our case? 

The answer is no. 
In our case, this procedure would mean to declare the radius
$r_{0,n}=1.881\,{\rm fm}$ where the normal force exhibits a 
node to be the ``edge" of the system. This ``works" in the 
following sense. If we multiply (\ref{eq:thedifferentialequation})
by $r^3$, integrate over a finite integral $0\le r \le R_n$,
and perform integrations by parts, we obtain \cite{Polyakov:2018zvc}
\be
       \int_0^{r}dr^\prime\;{r^\prime}^2p(r^\prime)=\frac{r^3}{3}\biggl(
       \frac23\,s(r)+p(r)\biggr)\,.
\ee
This means that the von Laue condition (\ref{eq:Laue}) 
could {\it also} be satisfied by integrating
over a {\it finite} interval from zero up to the node
of $\frac23\,s(r)+p(r)$ at $r_{0,n}=1.881\,{\rm fm}$.
From a mechanical stability point of view, we could be
happy about such a solution. From physical point of view,
we are not. While the effects of the short-range strong fields 
are practically negligible beyond $r_{0,n}=1.881\,{\rm fm}$
one cannot ignore the effect of the long-range Coulomb field, 
which ``communicates'' the presence of an electric charge.
We recall that the model \cite{BialynickiBirula:1993ce} was
designed with the specific purpose to have a mechanical model
of an electric charge --- which inevitably includes a correct
description (within Maxwell's equations) of its long-range
Coulomb potential. The ``truncation'' of the system at a finite 
value of $r$ is therefore unacceptable (in our case; for the 
macroscopic neutron stars it surely works).
We must seek a solution along different paths.

One possible resolution lies
in exploring the force concept in a classical system.
Notice that in our system, the matter (``continuous medium'') is 
described by the scalar density $\rho(r)$ localized within the radius 
$R=1.05\,{\rm fm}$. Thus, in the volume where the dust particles 
are present, the criterion (\ref{eq:positivity-norm-force}) 
{\it is} satisfied. 
The violation of (\ref{eq:positivity-norm-force}) occurs where
no matter is present. Hence, it does not affect the mechanical 
stability of the medium.
This argument cannot be applied to quantum field theoretical systems
where the distributions of ``matter'' and ``field energy'' cannot in 
general be distinguished \cite{Goeke:2007fp}. Only in classical systems, 
such as our model, is such a distinction unambiguous.

Thus, a possible resolution of the issue is that no violation of 
(\ref{eq:positivity-norm-force}) occurs in our model, because no matter 
is present at the point where the normal forces become negative. Hence, 
the node in the normal force causes no mechanical instability, and the 
classical proton solution is consistent.

Finally, we comment on the mean square radius of energy density 
$\la r_E^2\ra$ and mechanical mean square radius $\la r_{\rm mech}^2\ra$. 
These radii ``measure'' the extent of the spatial distributions of energy 
density and normal forces, which include the contributions of the fields,
and diverge due to the long-range Coulomb field. 
In this classical model, the ``proton size'' is associated with the 
localized distribution of matter. The matter particles carry electric
charge, and we already saw that system has a well-defined finite 
charge radius which numerically has the right order of magnitude,
see Sec.~\ref{Sec-2:model}.

\subsection{Regularized result \boldmath $D_{\rm reg}$ for the $D$-term}
\label{Subsec:4d-Dreg}

In this section, we address the question of whether the model
can make a prediction for the $D$-term. Taken literally, the
expressions (\ref{eq:Ds}) and (\ref{eq:Dp}) for the $D$-term
diverge, but one may try to regularize them. In general, 
regularization is not unique, and one must define a
``regularization prescription". Let us stress we are
talking here about the small contribution of the Coulomb
potential in Figs.~\ref{fig3}b and \ref{fig3}c, which spoil
the convergence of the integrals.

In our case, one can use the following procedure to obtain a 
finite result for the $D$-term. In order to motivate this
procedure, we notice that in a theory, where (a) the 
EMT is conserved and (b) the integrals defining $D_p$ and $D_s$
in Eqs.~(\ref{eq:Ds},~\ref{eq:Dp}) exist, one can compute the 
$D$-term in terms of an arbitrary linear combination 
of $D_p$ and $D_s$ as follows
\be
      D(\zeta) = \zeta\,D_p+(1-\zeta)\,D_s\,. \label{eq:Dreg}
\ee
Notice that under these conditions, the same result, 
$D(\zeta)=D$, is obtained for any value $\zeta$ which follows
simply from the equivalence of the expressions for $D_s$ and $D_p$. 
In our case, the condition (a) is of course satisfied, but (b) is not. 
As a consequence, the expression $D(\zeta)$ is divergent 
for all $\zeta$ except for one value $\zeta=\zeta_{\rm reg}$,
which can be chosen such that $D_{\rm reg}=D(\zeta_{\rm reg})$ 
is finite. This value of $\zeta_{\rm reg}$ depends on the number 
of dimensions $n=3$ and the power $N=4$ in the long-distance
asymptotic $s(r)=a_s/r^N$ and $p(r)=a_p/r^N$, and can be 
determined as follows.\footnote{%
    Notice that the integrals (\ref{eq:Ds},~\ref{eq:Dp}) 
    diverge if $N<n+2$, which is the case here since $N=4$ 
    and $n=3$. We remark that we left the dimensionality of space,
    $n=3$, general in Eqs.~(\ref{eq:thedifferentialequation}) and 
    (\ref{eq:Ds},~\ref{eq:Dp}) for the purposes of this 
    section.}
The coefficients $a_s$ and $a_p$ are not independent, 
but related to each other, as $a_p/a_s = - (n-1)(N-n)/(nN)$ due 
to Eq.~(\ref{eq:thedifferentialequation}). Thus, in the linear
combination, $s(r)/a_p+p(r)/a_s$, the long-range Coulomb tail cancels 
out. This is the only linear combination which can give a convergent 
result in Eq.~(\ref{eq:Dreg}). Considering the prefactors in the 
definitions  (\ref{eq:Ds},~\ref{eq:Dp}) of $D_s$ and $D_p$, the
required value of $\zeta_{\rm reg}$ is\footnote{For 
    completeness we remark that $N$ can be related to $n$
    by formulating the Maxwell's equations for a general number of
    dimensions. The ``area element'' is 
    $d\vec{a} = \vec{e}_r r^{n-1}\di\Omega_n$ in $n$-dimensional
    space. Then the Gauss law $\oint\vec{E}\,d\vec{a} = Q$ implies for
    the electric field of a localized charge distribution 
    $|\vec{E}|\propto 1/r^{n-1}$, and the Coulomb
    potential $A_0(r)\propto 1/r^{n-2}$. The long-distance
    behavior of EMT densities is determined by $A_0^\prime(r)^2$
    and given by $1/r^{2(n-1)}$. Hence $N=2(n-1)$. 
    }
\be
      \zeta_{\rm reg} = \frac{2N}{n(n+2-N)}\,.
\ee 
This is $\zeta_{\rm reg} = \frac83$ in our case for $n=3$ dimensions 
and $N=4$. The regularized $D$-term obtained in this way is finite, 
negative, expressed in terms of $p(r)$ and $s(r)$ as follows,
and numerically given by
\be\label{eq:D_reg}
    D_{\rm reg} = D(\zeta_{\rm reg}) = 
    M \int d^3r\;r^2\,\frac49\biggl[6p(r)+s(r)\biggr] = - 0.317\,(\hbar c)^2.
\ee
Numerically, this is about an order of magnitude smaller than the $D$-term
in the chiral quark soliton model \cite{Goeke:2007fp}. This is understandable 
considering the $D$-term encodes information on internal forces.
In the classical model, we deal with ``residual nuclear forces'' 
which are about an order of magnitude smaller than the forces 
among quarks in the chiral quark soliton model of Ref.~\cite{Goeke:2007fp}.

Our regularization method is distinguished because it removes the 
divergences from $D_p$ and $D_s$ in an efficient and ``minimalistic'' way: 
the long-range QED contribution exactly cancels out in the 
linear combination $[6p(r)+s(r)]$ in (\ref{eq:D_reg}) and we 
introduce no ``regulator dependence.'' It it is also a suitable
regularization because it preserves the negative sign of the
$D$-term observed so far in all theoretical studies. 

Another interesting feature of this regularization prescription is
that it removes the Coulomb contribution not only in the outer 
region $r>R$, which is required to make the $D$-term finite. 
Interestingly, in the linear combination of $p(r)$ and $s(r)$ in the
integrand in Eq.~(\ref{eq:D_reg}), the contribution of electrostatic 
forces also cancel out exactly  in the inner region $r<R$. In other 
words, $D_{\rm reg}$ receives no contribution from the 
electric forces at all. Notice that this concerns only the 
Coulomb contribution to $D_{\rm reg}$ in this particular
regularization scheme. The electromagnetic contribution to 
the budget of the internal forces is well-illustrated by the
von Laue condition in Eq.~(\ref{eq:Laue-in-detail}), where the
numerical contribution of the Coulomb field is small, but 
indispensable to prevent the collapse of the proton in the 
classical model.

However, our result (\ref{eq:D_reg}) is not unique, because in 
principle one could use other methods to regularize the divergences.
It would be interesting to see whether other suitable regularization 
methods can be defined, and investigate the effects of the regularization 
scheme dependence on the $D$-term. This will be left to future studies.

\newpage
\section{Model-independent insights, and the form factor \boldmath $D(t)$}
\label{Sec-5:model-independent-insights}

In this section, we put our findings in a wider context and show that the 
observed  long-distance properties of EMT densities are model independent. 
We discuss possible implications for theoretical and experimental studies 
of the EMT form factors, and especially the $D$-term form factor.

\subsection{EMT densities in QED at long-distances}

Our results for EMT densities are certainly model-dependent 
in the region up to $r\lesssim\,$2--3$\,{\rm fm}$, where strong forces 
dominate. The strong forces are modelled in a specific way in our
approach. One could
use a different model for strong forces and would obtain different results 
in the region $r\lesssim\,$2--3$\,{\rm fm}$. Independently, of the chosen 
model, the strong forces are short-ranged and their effects are faded 
out in the region $r\gtrsim3\,{\rm fm}$. This region is
governed by the long-range 
electromagnetic force.\footnote{\label{Footnote-large-r-chiral}%
       The point where electromagnetic force 
       becomes dominant, in our work $r\gtrsim3\,{\rm fm}$, is model 
       dependent. In QCD, the long distance-behavior of  nucleon EMT 
       densities is dictated by spontaneous chiral symmetry breaking 
       and the emergence of Goldstone bosons, pions in SU(2) flavor case,
       whose contributions to $s(r)$ and $p(r)$ decay like 
       $\frac{1}{r^4}\exp(-2m_\pi r)$ for pion mass $m_\pi\neq 0$, and are 
       proportional to $\frac{1}{r^6}$ in the chiral limit \cite{Goeke:2007fp}.
       At large enough distances, the contributions of electromagentic 
       forces dominate the EMT densities in any case.}

In other words, if one were to solve QCD and QED exactly, e.g.\ in 
a lattice calculation \cite{Hayakawa:2008an,deDivitiis:2013xla,
Borsanyi:2014jba,Endres:2015gda,Feng:2018qpx,Feng:2019geu}, one would 
recover the same results for the EMT densities at long distances 
$r \gg 3\,{\rm fm}$. The reason for that is obvious. 
The $\frac1r$-behavior of the classical
Coulomb potential is a consequence of the masslessness of the photon in QED.
Indeed, the long-distance part of the EMT densities in QED was determined from
1-loop  calculations in \cite{Donoghue:2001qc} and is given by
\be\label{eq:EMT-QED}
       T^{00}_{\rm QED}(\vec{r})=\frac{\alpha\hbar c}{8\pi r^4}+\dots\;,\quad
       T^{ij}_{\rm QED}(\vec{r}) = -\,\frac{\alpha\hbar c}{4\pi r^4}\,
       \biggl(e_r^ie_r^j-\frac12\,\delta_{ij}\biggr) + \dots
\ee
where the parenthesis denotes more strongly suppressed terms.
To this order, the results in Eq.~(\ref{eq:EMT-QED}) 
are the same for spin-0 and spin-$\frac12$ particles (while the components 
$T^{0k}_{\rm QED}(\vec{r})$ obviously depend on the spin) \cite{Donoghue:2001qc}.
The result (\ref{eq:EMT-QED}) can be traced back to non-analytic terms 
in the EMT form factors at small $t$ which arise from the 
masslessness of the photon \cite{Donoghue:2001qc}. The QED long-distance 
contributions to the EMT densities (\ref{eq:EMT-QED}) coincide exactly 
with our results (\ref{eq:long-distance}). This is not a coincidence,
but due to the fact that QED must reproduce the classical Maxwell theory 
at long distances \cite{Donoghue:2001qc}.

As we have seen, the long-distance behavior 
(\ref{eq:long-distance},~\ref{eq:EMT-QED}) implies a divergent $D$-term as 
well as the divergence of mean square radii associated with EMT densities.
These divergences are also model-independent results. This can be seen 
without involving the notion of EMT densities \cite{Donoghue:2001qc}.
For that we have to investigate the EMT form factors.

\subsection{The \boldmath $D$-term form factor}

EMT densities can be computed directly in classical models 
but not in quantum field theory\footnote{%
     One exception are quantum field theoretical models based on the 
     limit of a large number of colors $N_c$ where the nucleon structure 
     is described in terms of a mean field \cite{Witten:1979kh}, like in 
     chiral quark soliton or Skyrme model \cite{Goeke:2007fp}. 
     In the large-$N_c$ limit, the 3-dimensional EMT density formalism for
     baryons is exact \cite{Polyakov:2018zvc}. 
     Form factors also have an interpretation in terms of 2-dimensional 
     EMT densities, which is exact for all hadrons and valid for any $N_c$
     \cite{Burkardt:2000za,Burkardt:2002hr}, cf.\ Ref.~\cite{Lorce:2018egm}
     specifically for the case of EMT densities. \label{footnote-2}}
where all one can do is to evaluate matrix elements of the EMT operator
$\hat{T}^{\mu\nu}$. The information content of these matrix elements is 
described in terms of form factors which are defined in the case of the
nucleon as
\ba
    \la p^\prime,s^\prime| \hat T^{\mu\nu} |p,s\rangle
    = \bar u(p^\prime,s^\prime)\biggl[
      A(t)\,\frac{P^\mu P^\nu}{M}
    + J(t)\ \frac{i\,(P^\mu\sigma^{\nu\rho}+P^{\nu}\sigma^{\mu\rho})\Delta_\rho}{2M}
    + D(t)\,\frac{\Delta^\mu\Delta^\nu-g^{\mu\nu}\Delta^2}{4M}\biggr]u(p,s)
    \label{Eq:EMT-FFs-spin-12-alternative} \ea
where $P= \frac12(p^\prime + p)$, $\Delta = p^\prime-p$, $t=\Delta^2$, 
and the spinors are normalized as $\bar u(p,s)\, u(p,s) =2M$.
The EMT densities can be inferred indirectly from the form factors through an 
interpretation of the 3-dimensional Fourier transforms of the form factors. 
This interpretation is justified if the size of the particle is much 
larger than its Compton wavelength (which is the case for the proton to a 
good approximation) and is applicable for $r\gtrsim \lambda_c$ where 
$\lambda_c = \hbar/(Mc)\approx 0.2\,{\rm fm}$ denotes the Compton 
wave-length of the proton \cite{Polyakov:2018zvc}. 

\begin{figure}[b!]
  \begin{center}
    \includegraphics[width=0.25\textwidth]{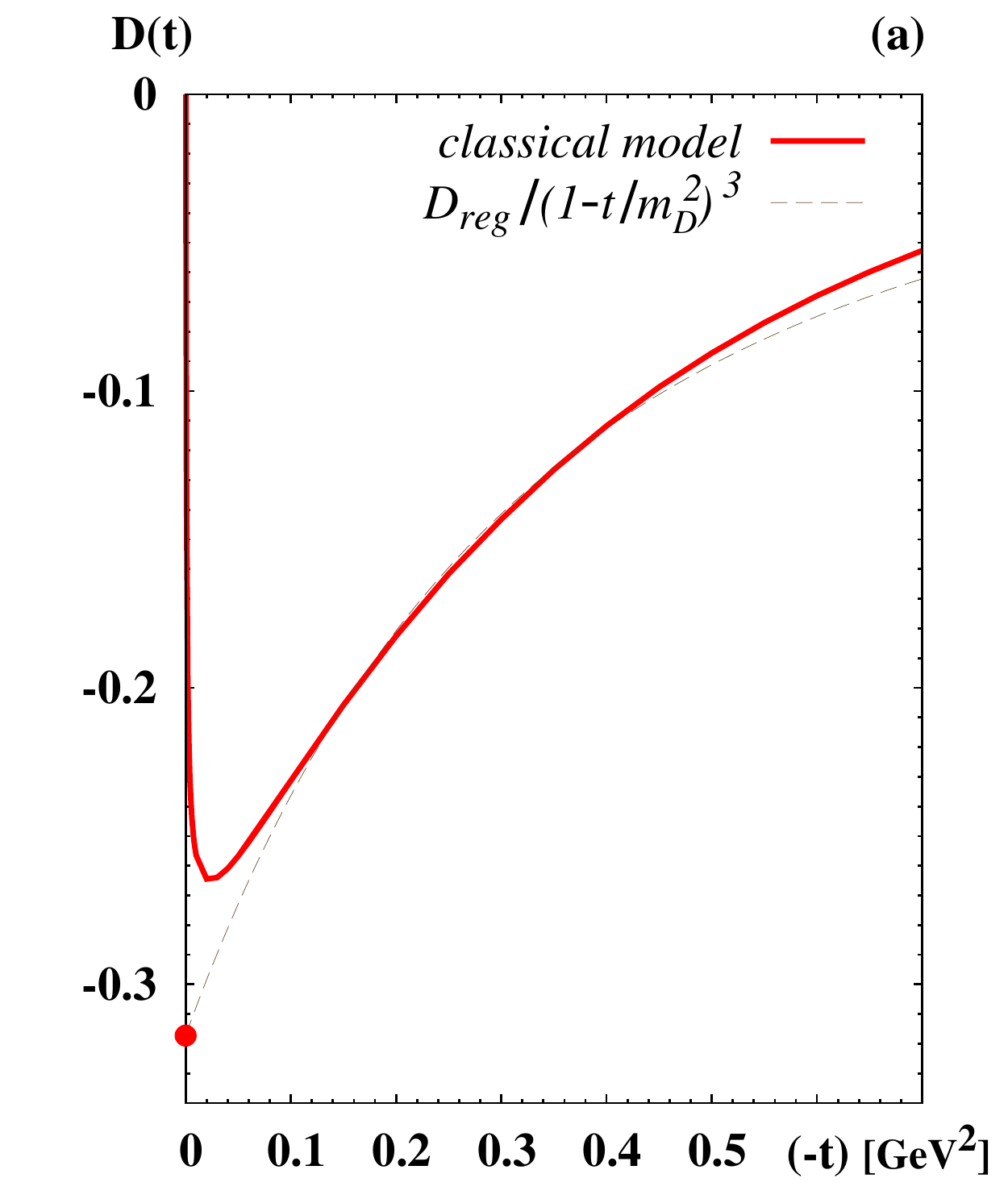}
    \includegraphics[width=0.25\textwidth]{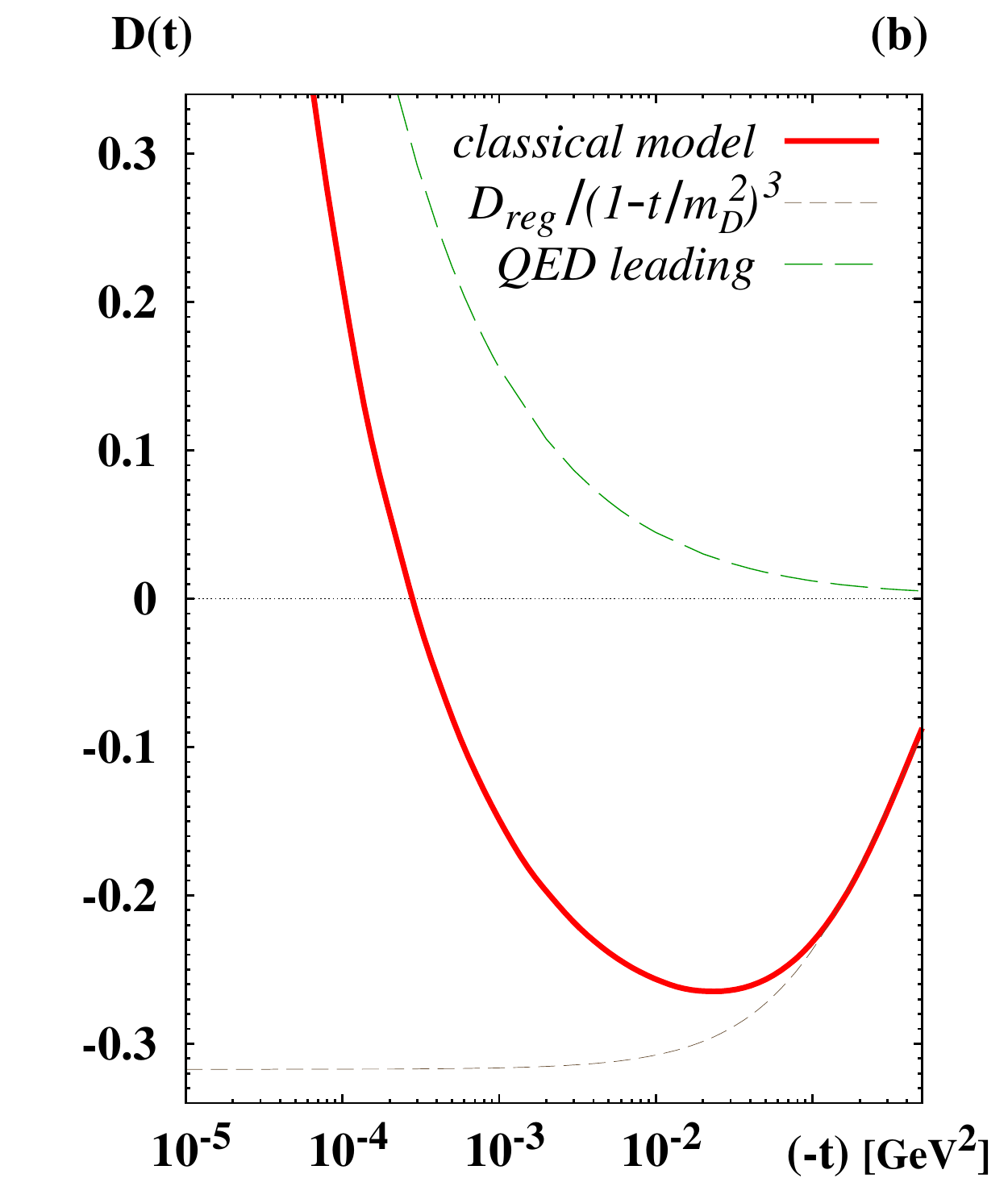}
    \includegraphics[width=0.25\textwidth]{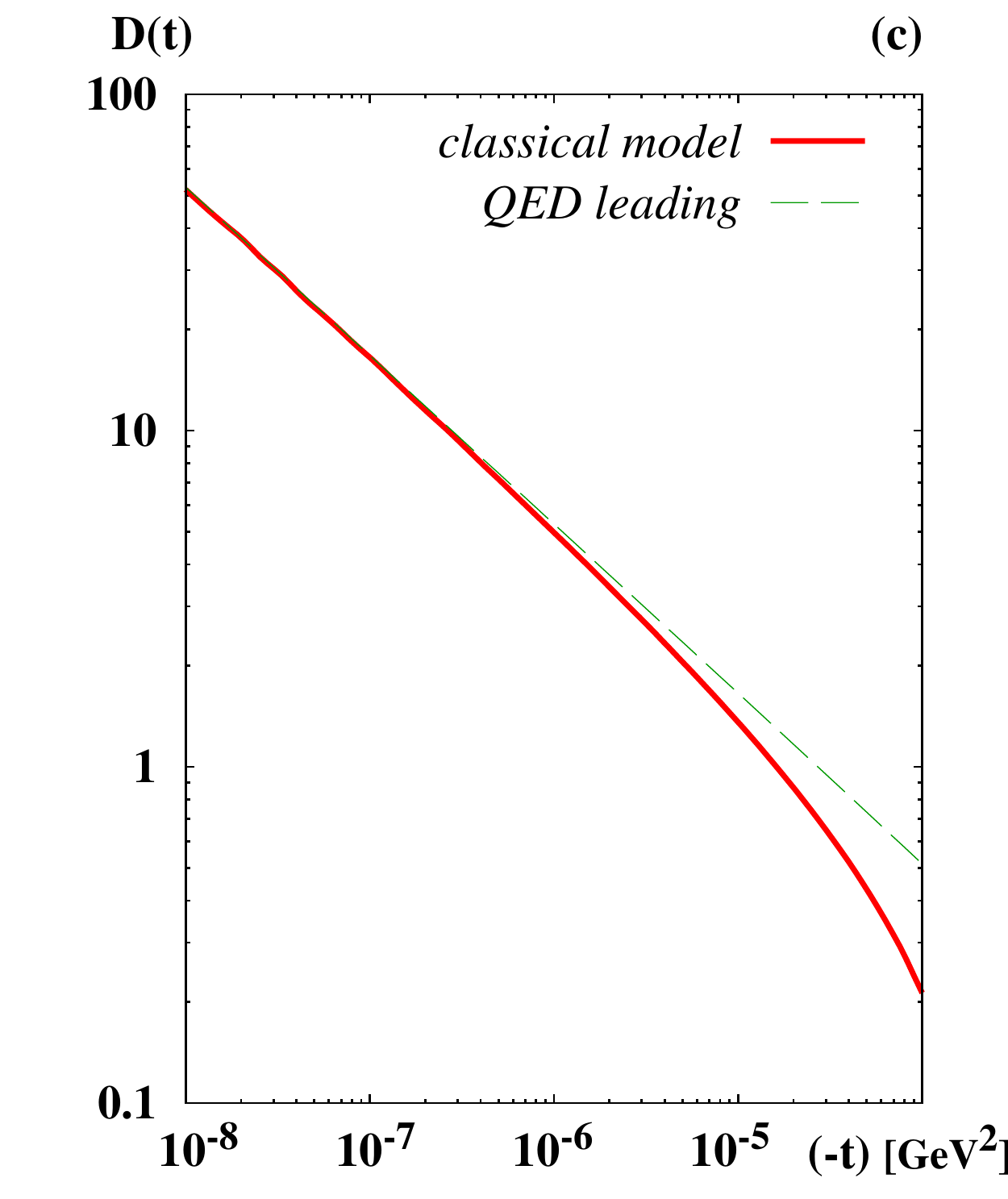}
  \end{center}
  \caption{\label{fig4}
      The form factor $D(t)$ in the classical model as function of $t$. 
      (a) Region of $0 \le (-t) \le 0.7\,{\rm GeV}^2$, where we find
      a typical $t$-dependence which can be approximated by an analytic 
      quadrupole form $D(t)_{\rm approx}\approx D_{\rm reg}/(1-t/m_D^2)^3$, with 
      $m_D=0.985\,{\rm GeV}$ for $0.1\lesssim(-t)\lesssim0.5\,{\rm GeV}^2$.
      The point marks the regularized value $D_{\rm reg}$ of the $D$-term from 
      Eq.~(\ref{eq:Dreg}). 
      (b) Transition region $10^{-5} \le (-t) \lesssim 10^{-1}\,{\rm GeV}^2$ 
      where $D(t)$ changes sign and starts to approach prediction from
      ``QED leading'' non-analytic terms.
      (c) The region $10^{-8} \le (-t) \le 10^{-4}\,{\rm GeV}^2$ of 
      ``asymptotically small'' $t$. For $(-t)\le 10^{-6}\,{\rm GeV}^2$, 
      the classical model result coincides with the QED prediction
      (\ref{eq:D-term-QED}) for $D(t)$ which diverges like $1/\sqrt{-t}$.}
\end{figure}

The interpretation of the form factors in terms of densities is performed
in a frame where $t=-\vec{\Delta}^{\,2}$. If the form factor $D(t)$ is known,
one way to determine the stress tensor densities $s(r)$ and $p(r)$ is as
follows \cite{Polyakov:2018zvc}
\ba
       \widetilde{D}(r) &=& \int\!\frac{d^3\Delta}{(2\pi)^3}
       e^{-i\vec{\Delta}\cdot\vec{r}} \, D(-\vec{\Delta}^{\,2}) \, ,
       \label{Eq:Dtilde}\\
       s(r) &=& -\,\frac{1}{4M}\,r\,\frac{d\,}{dr}
       \biggl[\,\frac1r\,\frac{d\,}{dr}\,\widetilde{D}(r)\biggr]\,,
       \label{Eq:Dtilde-s}\\
       p(r) &=& \,\frac{1}{6M}\,\frac{1}{r^2}\;\frac{d\,}{dr}
       \biggl[\,r^2\,\frac{d\,}{dr}\widetilde{D}(r)\biggr]\,.
       \label{Eq:Dtilde-p}
\ea
Here we can proceed ``backwards'' and integrate 
Eqs.~(\ref{Eq:Dtilde-s},~\ref{Eq:Dtilde-p}) to obtain 
$\widetilde{D}(r)$. This is most conveniently done by integrating
 the expressions (\ref{Eq:Dtilde-s},~\ref{Eq:Dtilde-p}) twice over 
the radial distances from $r$ to infinity,
where all densities vanish. Starting from Eq.~(\ref{Eq:Dtilde-s})
or (\ref{Eq:Dtilde-p})  respectively yields the same result for
$\widetilde{D}(r)$ which serves as a test for the calculation. 
With the result for $\widetilde{D}(r)$, one can invert the Fourier 
transform (\ref{Eq:Dtilde}) to compute $D(t)$. The result is shown 
in Fig.~\ref{fig4}.

The Fig.~\ref{fig4}a gives an overview of $D(t)$.
In the region $0.1 \lesssim (-t) \lesssim 0.5\,{\rm GeV}^2$,
we observe a shape typical for hadronic form factors, which can
be approximated by a quadrupole form 
$D(t)_{\rm approx}\approx D_{\rm reg}/(1-t/m_D^2)^3$, with $m_D=0.985\,{\rm GeV}$.
At smaller $(-t)$, the QED long-distance effects become noticeable. 
At significantly larger $(-t)$, the correct description of form factors 
requires short-distance properties of QCD which are not present in the model. 
As the $t=0$ intercept of $D(t)_{\rm approx}$, we have chosen the regularized 
value $D_{\rm reg}$ from Eq.~(\ref{eq:Dreg}), which is 
indicated in Fig.~4a (one could also choose a slightly different value).
The quadrupole shape of $D(t)_{\rm approx}$ is suggested by large-$t$ 
QCD counting rules \cite{Polyakov:2018zvc} (but one could also choose 
other shapes). For $t>0.1\,{\rm GeV}^2$,
the classical model result for $D(t)$ is in good qualitative
agreement with more realistic models \cite{Goeke:2007fp} but about an order 
of magnitude smaller which is expected, cf.\ discussion below
Eq.~(\ref{eq:D_reg}). 
For $(-t) < 0.1\,{\rm GeV}^2$, the long-distance QED effects
start to become important, and cause the form factor $D(t)$ to
change sign at $t=-2.8\times 10^{-4}\,{\rm GeV}^2$. This 
``transition region'' is shown in Fig.~\ref{fig4}b. 
In this region, we can start to compare our results to the predictions 
for $D(t)$ due to the QED leading non-analytic terms \cite{Donoghue:2001qc}. 

\subsection{\boldmath Comparison to the leading non-analytic QED 
contributions to $D(t)$}

The leading non-analytic QED terms in the small-$t$ behavior of EMT 
form factors were derived in \cite{Donoghue:2001qc} (where the 
notation $q^2=t$, $F_1(q^2)=A(t)$, $F_2(q^2)=2J(t)$, $F_3(q^2)=\frac14D(t)$
was used). 
The derivation of the long-distance QED contribution to the stress tensor
quoted earlier in Eq.~(\ref{eq:EMT-QED}) was part of a calculation of QED 
one-loop corrections to the gravitational metric of charged particles with 
spin~0 and spin~$\frac12$ \cite{Donoghue:2001qc}. 
The calculations were performed using effective field theory techniques.
For a charged spin-$\frac12$ fermion, the small-$t$ behavior of $D(t)$ 
due to QED-effects is \cite{Donoghue:2001qc}
\be\label{eq:D-term-QED}
      D(t) = \frac{\alpha}{\pi}\biggl(-\,\frac{11}{18}
           + \frac{\pi^2 M}{4\sqrt{-t}}+\frac23\,\log\frac{(-t)}{M^2}\biggr)
           + \dots \,,
\ee
where the dots indicate terms which are finite as $t\to0$. Notice 
that $D(t)$ is multiplied by $(\Delta^\mu\Delta^\nu-g^{\mu\nu}\Delta^2)$
in (\ref{Eq:EMT-FFs-spin-12-alternative}). Therefore, the matrix element
$ \la p^\prime,s^\prime| \hat T_{\mu\nu} |p,s\rangle$  has a 
well-defined limit $t\to0$, but the form factor $D(t)$ does not. 
The $D$-term given by $D=D(0)=\lim_{t\to 0}D(t)$ is divergent. 
The result for the EMT densities (\ref{eq:EMT-QED})
was obtained from Eq.~(\ref{eq:D-term-QED}) by means of a Fourier 
transform. The EMT determines the metric through the Einstein 
equation, and from the long-distance QED contribution to the EMT
densities (\ref{eq:EMT-QED}), it is possible to reproduce the classical 
non-linear terms of the Reissner-Nordstr\"om metric for a non-spinning 
charge or Kerr-Newman metric for a spinning charge 
\cite{Donoghue:2001qc}.

It is important to stress that the metric in general relativity is an
inherently classical concept. The deeper reason why it is possible to 
determine quantum corrections to the metric lies in the massless nature 
of the photon, which causes long-distance effects much stronger than the 
gravitational effects, provided $\alpha\gg G\,M^2/(\hbar c)=M^2/M_{\rm Planck}^2$,
where $G$ denotes the gravitational constant and $M_{\rm Planck}$ is the 
Planck mass. Under this condition, quantum gravity corrections can be 
neglected, and gravity can be treated as a classical theory described
in terms of a metric (in our case quantum gravity effects can safely 
be neglected: the proton mass $M$ is 19 orders of magnitude smaller 
than $M_{\rm Planck}=1.2\times 10^{19}\,{\rm GeV}$). Besides the photon
\cite{Donoghue:2001qc}, graviton effects can also be studied in this way
\cite{Donoghue:1993eb,Khriplovich:2002bt,BjerrumBohr:2002kt,Khriplovich:2004cx}.

The QED result (\ref{eq:D-term-QED}) is shown in Fig.~\ref{fig4}b
with the label ``QED leading.''
After the sign change, the model result for $D(t)$ starts to slowly 
approach the QED result (\ref{eq:D-term-QED}). For ``asymptotically small'' $t$ 
in the region below $(-t)\le 10^{-6}\,{\rm GeV}^2$, the classical model result 
for $D(t)$ practically coincides with the QED result (\ref{eq:D-term-QED}),
see Fig.~\ref{fig4}c. In particular, the model result for $D(t)$ 
diverges like $1/\sqrt{-t}$ for small $(-t)$.

For completeness, we remark that the form factors $A(t)$ and $J(t)$ 
must satisfy the constraints $A(0)=1$ and $J(0)=\frac12$ 
\cite{Lowdon:2017idv,Cotogno:2019xcl,Lorce:2020bsg} and do so
of course despite the presence of long-range QED corrections.
In the case of these form factors, the leading non-analytic terms
are of the type $\sqrt{-t}$ and $(-t)\log(-t)$ and well-behaving
for $t\to0$. But the derivatives of these form factors with respect to
$t$ diverge in the limit $t\to0$. For instance, in the case of $A(t)$, 
this implies the divergence of the ``gravitational mean square radius'' 
\cite{Donoghue:2001qc} is defined as $\la r^2_{\rm grav}\ra=-6A^\prime(0)$. 
The mean square radius of the energy density is also related to
$A^\prime(0)$ \cite{Polyakov:2018zvc} and divergent due to 
the long-range QED effects, see above.

\subsection{Consequences for calculations and measurements of \boldmath $D(t)$}

In theoretical studies of the hadron structure, electromagnetic effects 
can often be neglected to a good approximation.
But in the experiment, one certainly cannot neglect the electric charge 
of the proton and other electromagnetic effects. 
Even though not straightforward \cite{Polyakov:2018zvc}, 
the form factor $D(t)$ of the proton can be extracted from analyses 
of hard exclusive reactions like deeply virtual Compton scattering 
\cite{Mueller:1998fv,Ji:1998pc} using dispersion relation methods
\cite{Teryaev:2005uj,Anikin:2007yh,Diehl:2007jb,Radyushkin:2011dh}
and first attempts were reported \cite{Nature,Kumericki:2019ddg}. 
Can the divergent behavior of $D(t)$ due to QED effects 
(\ref{eq:D-term-QED}) be experimentally observed?

It is important to stress that the QED contribution to $D(t)$ in 
Eq.~(\ref{eq:D-term-QED}) starts to become noticeable in our model 
only for $(-t)\ll 0.1\,{\rm GeV}^2$. In more realistic models, the 
contribution of strong forces to $D(t)$ is an order of magnitude 
larger, implying that the transition region where $D(t)$ changes 
sign, cf.\ Fig.~\ref{fig4}b, sets in at even lower values of $(-t)$.
In addition, in the case of deeply virtual Compton scattering, 
it is necessary to consider higher order QCD corrections.
When extracting information from electromagnetic processes, one must 
also consider QED radiative corrections, which are generically 
of the same order of magnitude as the effect (\ref{eq:D-term-QED}). 
It is by no means obvious whether the result (\ref{eq:D-term-QED}) 
can be disentangled from radiative corrections. Even if it can,
it remains to be seen whether such corrections can be determined 
with sufficient precision to observe (\ref{eq:D-term-QED}). Thus,
from a practical point of view, one may never be able to reach the 
region of small enough $(-t)$, cf.\ Figs.~\ref{fig4}b and \ref{fig4}c, 
and sufficient precision to observe QED effects like (\ref{eq:D-term-QED}). 

But from a theoretical point of view, it is legitimate to ask how 
to calculate the $D$-term in a system with long-range forces. 
We do not know a definite answer to this question, though our work 
indicates a possible solution, namely to apply a regularization scheme. 
At first glance, it may appear unusual to invoke regularization in  classical 
calculations. But the deeper reason why the $D$-term diverges is rooted in the 
masslessness of the photon, and hence related to infrared divergences in QED.
The regularization prescription proposed in Sec.~\ref{Subsec:4d-Dreg}
is acceptable because: 
(i)  it gives a finite result,
(ii) preserves the negative value of the $D$-term 
in accordance with other theoretical studies, and 
(iii) the obtained numerical value (\ref{eq:Dreg}) is useful to practically 
approximate $D(t)$ at finite $(-t)>0.1\,{\rm GeV}^2$, see Fig.~\ref{fig4}a.
Thus, working with a such a regularization scheme is one practical way out.
The regularization method of Sec.~\ref{Subsec:4d-Dreg} works in our
classical calculation. In perturbative QCD and QED calculations of the deeply 
virtual Compton scattering process, of course other ``schemes'' are invoked, 
and in nonperturbative lattice QCD calculations with included QED effects, 
one can use yet other regularization methods 
\cite{Hayakawa:2008an,Endres:2015gda,Feng:2018qpx}. More
theoretical work is required to compare results obtained in
different schemes. These aspects go beyond the scope of this work.

\section{Conclusions}
\label{Sec-6:conclusions}

Prior EMT studies focused on applications to hadronic physics, and
considered mainly strongly interacting systems with short-range forces.
Long-range forces were not included. In systems governed by short-range 
forces, the $D$-term was always found to be well-defined, finite and negative.
In this work, we have presented a study in a system where in addition to 
(strong) short-range forces,  (electromagnetic) long-range forces are also
present. We have encountered several interesting features not observed in 
prior studies of systems with short-range forces. The most interesting 
observation is that, when long-range forces are included, the $D$-term 
is no longer well-defined, and diverges. 

For our study, we employed the classical proton model of 
Bia\l ynicki-Birula \cite{BialynickiBirula:1993ce} which is of interest
for its own~sake. To the best of our knowledge, it is the first fully 
consistent classical model of an extended charged particle where the 
Poincar\'e stresses are generated dynamically in a local, relativistic, 
classical field theory. The model exhibits short-range strong forces,
which are modelled after nuclear forces, and the electromagnetic 
long-range interaction \cite{BialynickiBirula:1993ce}.
The two crucial aspects for our study are the consistent description of a 
stable particle, and correct description of the long-range electromagnetic 
effects. The model of Ref.~\cite{BialynickiBirula:1993ce} satisfies both.
The classical aspect of the model is an advantage in the sense that 
it allows us to concentrate on the effects of long-range forces undistracted 
by technical difficulties, which are inevitable in studies of more realistic, 
strongly interacting quantum systems. 

In the region below $r\lesssim2\,$fm, the classical proton model yields 
results for the energy density $T_{00}(r)$, shear force $s(r)$, and 
pressure $p(r)$ in good qualitative agreement with more realistic models 
like the chiral quark soliton or Skyrme model, 
except that $s(r)$ and $p(r)$ are about an order of magnitude smaller. 
This is of course to be expected in a model where the internal forces 
are modelled after the ``residual nuclear forces.'' Otherwise, 
in this $r$-region, the
classical proton model is in line with results from short-range systems.

The situation is different for $r\gtrsim2$--$3\,$fm, where the strong
forces are faded out, and $T_{00}(r)$, $s(r)$, $p(r)$ exhibit 
tails proportional to $\frac1{r^4}$ due to the long-range Coulomb field. 
This introduces new features, e.g.\ $s(r)$ and $p(r)$ show (additional) 
nodes and opposite signs at large-$r$ as compared to systems with 
short-range forces. 
Another consequence is the divergence of the mean square radius 
of the energy density and the mechanical radius. These radii measure the 
spatial extensions of the energy density $T_{00}(r)$ and normal force 
$\frac23\,s(r)+p(r)$, which include contributions from~fields. 
Due to the long-range of the Coulomb field, the size of the 
system, as measured by these radii, is consequently infinite. 
In contrast, the electric mean square radius is finite 
(and numerically of the right size) \cite{BialynickiBirula:1993ce},
as it measures the spatial extension of the electric charge distribution 
tight to the localized matter distribution in the model. 

The most interesting new feature is that the $D$-term of a charged
particle is divergent. Technically, this happens because the $D$-term 
is given in terms of integrals over $s(r)$ or $p(r)$ which diverge 
due to the  $\frac1{r^4}$-tails of these densities. 
We proposed a regularization scheme which yields a finite, negative, 
and numerically reasonable value for a system with internal interactions 
of the strength of ``residual nuclear forces.'' We computed the 
form factor $D(t)$ which, for $(-t)>0.1\,{\rm GeV}^2$, is negative 
and shows a shape typical for hadronic form factors.
But in the region $(-t)\ll 0.1\,{\rm GeV}^2$, the form factor
changes sign, becomes positive and diverges as $t\to0$.

The observed long-distance properties of EMT densities and the related 
small-$t$ divergence of $D(t)$ are model-indepedent results. Both have 
been derived in Ref.~\cite{Donoghue:2001qc} from QED diagrams. 
In a recent study of $Q$-balls carrying an electric charge, the same 
long-range tails were found as in our work \cite{Loginov:2020xoj}. 
The deeper reason for the emergence of these EMT properties can be 
traced back to the masslessness of the photon \cite{Donoghue:2001qc}, 
which is reflected in the classical Maxwell's equations. Consequently,
the EMT long-distance properties must be correctly reproduced in every 
system (classical, quantum mechanical, quantum field theoretical) 
where the electromagnetic interaction is correctly described.

Other EMT form factors are also affected by QED long-distance effects, 
but less strongly than $D(t)$ for two reasons. 
First, the proton EMT form factors $A(t)$ and $J(t)$ are constraint 
to satisfy $A(0)=1$ and $J(0)=\frac12$ 
\cite{Lowdon:2017idv,Cotogno:2019xcl,Lorce:2020bsg}. QED 
long-distance effects must preserve these constraints, though
they can (and do) affect the derivatives of these form~factors 
(making them infinitely steep at $t=0$ which in turn is connected to 
the divergence of the related mean square radii). 
The value of $D(t)$ at $t=0$ is in general not constrained by any 
principle, and can therefore show more variation than other form factors. 
Second, the $D$-term is intrinsically related to the internal forces and 
the dynamics in a system. As such, it is the particle property which
exhibits by far the strongest sensitivity to variations in the system. 
It is therefore not unexpected that $D(t)$ shows the most pronounced 
effects when long-range forces are included.

The long-distance QED effects on $D(t)$ become noticeable at such small 
$(-t)\ll 0.1\,{\rm GeV}^2$  that it is not clear whether they are, 
even in principle, measurable. More theoretical work is required to 
clarify this important point. Our results are very interesting from a
theoretical point of view, and raise the question of how to define the
$D$-term in a system with long-range forces. Another well-known
long-range force is gravity. It would be very interesting
to perform a fully consistent computation of EMT properties 
including general relativity. Solutions of the Einstein equation for 
a perfect charged fluids exist, see e.g.\ \cite{Ivanov:2002jy}, but 
they typically make assumptions about the density or the equation of state.
A consistent treatment of the $D$-term requires to exactly solve the
dynamics of all involved fields, including the gravitational field.
As noted in \cite{Hudson:2017xug}, due to its sensitivity to 
the details of the involved interactions, the correct definition of 
the $D$-term may require the consideration of all forces, including 
QED and perhaps even gravity.

\newpage
\noindent{\bf Acknowledgments.} 
The authors wish to thank Luchang Jin and Maxim Polyakov for valuable 
discussions. This work was supported by the National Science Foundation 
under the Contract No.\ 1812423.

\end{document}